\documentclass[10pt]{article}
\usepackage{amsmath}
\usepackage{latexsym}
\usepackage{amssymb}
\usepackage{amsfonts}
\usepackage{amsthm}
\title{A MATHEMATICAL COMPARISON OF THE \\
SCHWARZSCHILD AND KERR METRICS}
\author{J.-F. Pommaret \\ CERMICS, Ecole des Ponts ParisTech, France \\
jean-francois.pommaret@wanadoo.fr  \\
 http://cermics.enpc.fr/$\sim$pommaret/home.html }
\date{  }
\begin{document}
\maketitle

\noindent
{\bf ABSTRACT}  \\

A few physicists have recently constructed the generating compatibility conditions (CC) of the Killing operator for the Minkowski (M) , Schwarzschild (S) and Kerr (K) metrics. They discovered second order CC, well known for M, but also third order CC for S and K. In a recent paper, we have studied the cases of M and S, without using specific technical tools such as Teukolski scalars or Killing-Yano tensors. However, even if S($m$) and K($m,a$) are depending on constant parameters in such a way that S $\rightarrow $ M when $m \rightarrow 0$ and K $\rightarrow$ S when  $ a \rightarrow 0$, the CC of S do not provide the CC of M when $m \rightarrow 0$ while the CC of K do not provide the CC of S when $a\rightarrow 0$. In this paper, using tricky motivating examples of operators with constant or variable parameters, we explain why the CC are depending on the choice of the parameters. In particular, the only purely intrinsic objects that can be defined, namely the extension modules, may change drastically. As the algebroid bracket is compatible with the {\it prolongation/projection} (PP) procedure, we provide for the first time all the CC for K in an intrinsic way, showing that they only depend on the underlying Killing algebras and that the role played by the Spencer operator is crucial. We get K$<$S$<$M with $2 < 4 < 10$ for the Killing algebras and explain why the formal search of the CC for M, S or K are strikingly different, even though each Spencer sequence is isomorphic to the tensor product of the Poincar\'{e} sequence for the exterior derivative by the corresponding Lie algebra.

\vspace{3cm}

\noindent
{\bf KEY WORDS}  \\

\noindent
Formal integrability; Involutivity; Compatibility condition; Janet sequence; Spencer sequence; Minkowski metric; Schwarzschild metric; Kerr metric.

\newpage

\noindent
{\bf 1) INTRODUCTION}  \\

In order to explain the type of problems we want to solve, let us start adding a constant parameter to the example provided by Macaulay in $1916$ that we have presented in a previous paper for other reasons ([14]). We first recall the following key definition:  \\

\noindent
{\bf DEFINITION 1.1}: A {\it system} of order $q$ on $E$ is an open vector subbundle $R_q\subseteq J_q(E)$ with prolongations 
${\rho}_r(R_q)=R_{q+r}=J_r(R_q)\cap J_{q+r}(E)\subseteq J_r(J_q(E))$ and symbols ${\rho}_r(g_q)=g_{q+r}=S_{q+r}T^*\otimes E \cap  R_{q+r}\subseteq  J_{q+r}(E)$ only depending on $g_q\subseteq S_qT^*\otimes E$. For $r,s\geq 0$, we denote by $R^{(s)}_{q+r}={\pi}^{q+r+s}_{q+r}(R_{q+r+s})\subseteq R_{q+r}$ the projection of $R_{q+r+s}$ on $R_{q+r}$, which is thus defined by more equations in general. The system $R_q$ is said to be {\it formally integrable} (FI) if we have $R^{(s)}_{q+r}=R_{q+r}, \forall r,s\geq 0$, that is if all the equations of order $q+r$ can be obtained by means of only $r$ prolongations. The system $R_q$ is said to be {\it involutive} if it is FI with an involutive symbol $g_q$. We shall simply denote by 
$\Theta=\{ f\in E \mid j_q(f) \in R_q\}$ the "set" of (formal) solutions. It is finally easy to prove that the Spencer operator $D:J_{q+1}(E) \rightarrow T^*\otimes  J_q(E)$ restricts to $D:R_{q+1} \rightarrow T^*\otimes R_q$. \\

The most difficult but also the most important theorem has been discovered by M. Janet in $1920$ ([7]) and presented by H. Goldschmidt in a modern setting in $1968$ ([5]). However, the first proof with examples is not intrinsic while the second, using the Spencer operator, is very technical and we have given a quite simpler  different proof in $1978$ ([7], also [9],[10]) that we shall use later on for studying the Killing equations for the Schwarzschild and Kerr metrics:   \\

\noindent
{\bf THEOREM 1.2}: If $R_q\subset J_q(E)$ is a system of order $q$ on $E$ such that its first prolongation $R_{q+1}\subset J_{q+1}(E)$ is a vector bundle while its symbol $g_{q+1}$ is also a vector bundle, then, if $g_q$ is $2$-acyclic, we have ${\rho}_r(R^{(1)}_q)=R^{(1)}_{q+r}$.  \\

\noindent
{\bf COROLLARY 1.3}: (PP {\it procedure}) If a system $R_q \subset J_q(E)$ is defined over a differential fiel $K$, then one can find integers $r,s \geq 0$ such that $R^{(s)}_{q+r}$ is formally integrable or even involutive.  \\

Starting with an arbitrary system $R_q\subset J_q(E)$, the main purpose of the next crucial example is to prove that the generating CC of the operator 
${\cal{D}}={\Phi}_0 \circ j_q : E \stackrel{j_q}{\longrightarrow}J_q(E) \stackrel{{\Phi}_0}{\longrightarrow } J_q(E)/R_q=F_0$, though they are of course fully determined by the first order CC of the final involutive system $R^{(s)}_{q+r}$ produced by the up/down PP procedure, are in general of order $r+s+1$ like the Riemann or Weyl operators, but may be of strictly lower order. \\

\noindent
{\bf MOTIVATING EXAMPLE 1.4} : With $m=1, n=3, q=2$, let us consider the second order linear system $R_2 \subset J_2(E)$ with $ dim (R_2)= 8$ and parametric jets $\{ y, y_1, y_2, y_3, y_{11}, y_{12}, y_{22}, y_{23}\}$, defined by the two inhomogeneous PD equations where $a$ is a constant parameter:  \\
\[  Py\equiv y_{33} = u, \hspace{2cm} Qy\equiv y_{13} + a\,y_2=v  \]
First of all we have to look for the symbol $g_2$ defined by the two linear equations $y_{33}=0, y_{13}=0$. The coordinate system is not $\delta$-regular and exchanging $x^1$ with $x^2$, we get the Janet board:  \\
\[ \left \{ \begin{array}{rcl}
  y_{33} &  =  & 0  \\
  y_{23}  &  =  &  0 
  \end{array} \right. \fbox{ $\begin{array}{lll}
  1 & 2 & 3  \\
  1 & 2 & \bullet
  \end{array} $ }   \]
 It follows that $g_2$ is involutive, thus $2$-acyclic and we obtain from the main theorem ${\rho}_r(R^{(1)}_2)= R^{(1)}_{2+r}$. However, $R^{(1)}_2 \subset R_2$ with a strict inclusion because $R^{(1)}_2$ with $dim(R^{(1)}_2)= 7$ is now defined by the 3 equations:  \\
 \[    y_{33}= u , \hspace{1cm}  a \, y_{23}=  v_3 - u_1, \hspace{1cm}  y_{13} +a \,y_2= v   \]
We may start again with $R^{(1)}_2$ and study its symbol $g^{(1)}_2 $ defined by the 3 linear equations with Janet tabular:  \\
\[  \left \{ \begin{array}{rcl}
  y_{33} &  =  & 0  \\
 a \, y_{23}  &  =  &  0 \\
  y_{13}  &  =  &  0
  \end{array} \right. \fbox{ $\begin{array}{lll}
  1 & 2 & 3  \\
  1 & 2 & \bullet  \\
  1  &  \bullet &  \bullet 
  \end{array} $ }   \]Since that moment, we have to consider the two possibilities:  \\
  
  \noindent 
$\bullet$ $a=0$:  The initial system becomes $y_{33}=u, y_{13}=v$ and has an involutive symbol. It is thus involutive because it is trivially FI 
as the left members are homogeneous with only one generating first order CC, namely $ u_3 - v_1=0$. We have $dim(g_{2+r})=4+r$ and the following commutative and exact diagrams:  \\
\[  \begin{array}{rcccccccl}
    &  0  &  &  0  &  &  0  &  & 0  &   \\
     &  \downarrow &  & \downarrow &  &  \downarrow  &  & \downarrow  &     \\
0 \rightarrow  &  g_3  & \rightarrow & S_3T^*\otimes E & \rightarrow &  T^*\otimes F_0 & \rightarrow &   F_1 &  \rightarrow 0 \\
&  \downarrow &  & \downarrow &  &  \downarrow  &  &  \parallel    &     \\
0 \rightarrow &  R_3 & \rightarrow  & J_3(E) & \rightarrow  &  J_1(F_0) &  \rightarrow &  F_1  &  \rightarrow 0  \\
    &  \downarrow &  & \downarrow &  &  \downarrow  &  & \downarrow   &     \\
0 \rightarrow &  R_2 & \rightarrow  & J_2(E) & \rightarrow  &  F_0 &  \rightarrow &  0  &  \\ 
   & \downarrow &  & \downarrow &  &  \downarrow  &  &   & \\
   & 0 &  &  0  && 0  && &
\end{array}   \]
\[  \begin{array}{rcccccccl}
    &  0  &  &  0  &  &  0  &  & 0  &   \\
     &  \downarrow &  & \downarrow &  &  \downarrow  &  & \downarrow  &     \\
0 \rightarrow  &  5  & \rightarrow &10 & \rightarrow & 6 & \rightarrow &  1 &  \rightarrow 0 \\
&  \downarrow &  & \downarrow &  &  \downarrow  &  &  \downarrow    &     \\
0 \rightarrow &  13 & \rightarrow  & 20  & \rightarrow  & 8  &  \rightarrow &  1  &  \rightarrow 0  \\
    &  \downarrow &  & \downarrow &  &  \downarrow  &  & \downarrow   &     \\
0 \rightarrow & 8 & \rightarrow  &  10   & \rightarrow  & 2  &  \rightarrow &  0  &  \\ 
   & \downarrow &  & \downarrow &  &  \downarrow  &  &   & \\
   & 0 &  &  0  &  & 0  &  &  &
\end{array}   \]

We have thus the Janet sequence:  \\
\[        0 \rightarrow \Theta \rightarrow E \underset 2{\stackrel{{\cal{D}}}{\longrightarrow}} F_0 \underset 1{\stackrel{{\cal{D}}_1}{\longrightarrow}} F_1 \rightarrow 0  \]
or, equivalently, the exact sequence of differential modules over $D=\mathbb{Q} [d_1,d_2,d_3]=\mathbb{Q}[d]$: \\
\[                      0 \rightarrow D\underset 1{\stackrel{}{ \longrightarrow}Ê} D^2 \underset 2 {\stackrel{}{\longrightarrow}Ê} D \stackrel{p}{\longrightarrow} M \rightarrow 0   \]
where $p$ is the canonical projection onto the residual differential module.  \\

\noindent
$\bullet$  $a\neq 0$: When the coefficients are in a differential field of constants, for example if $a\in \mathbb{Q}$ is invertible, we may choose $a=1$ like Macaulay ([14]). It follows that $g^{(1)}_2$ is still involutive but we have the strict inclusion $g^{(1)}_2\subset g_1$ and thus the strict inclusion $R^{(1)}_2 \subset R_2$ because $dim(R^{(1)}_2)=7<8$. We may thus continue the PP procedure and obtain the new strict inclusion $R^ {(2)}_2 \subset R^{(1)}_2$ because $dim(R^{(2}_2)=6$ as $R^{(2)}_2 $ is defined by the 4 equations with Janet tabular:   \\
\[  \left \{ \begin{array}{lcl}
  y_{33} &  =  & u \\
  y_{23}  &  =  &  v_3 - u_1   \\
  y_{22}  &  =  & v_2 -v_{13} + u_{11} \\
  y_{13} + y_2  &  =  &  v
  \end{array} \right. \fbox{ $\begin{array}{lll}
  1 & 2 & 3  \\
  1 & 2 & \bullet  \\
  1 & 2 & \bullet   \\
  1  &  \bullet &  \bullet 
  \end{array} $ }   \]
As $R^{(2)}_2$ is easily seen to be involutive, we achieve the PP procedure, obtaining the strict intrinsic inclusions and corresponding fiber dimensions:  \\
\[ R^{(2)}_2 \subset  R^{(1)}_2 \subset  R_2    \hspace{1cm}  \Leftrightarrow  \hspace{1cm}   6  <  7  <  8  \]
Finally, we have ${\rho}_r(R^{(2)}_2)= {\rho}_r((R^{(1)}_2)^{(1)})=({\rho}_r(R^{(1)}_2))^{(1)}=(R^{(1)}_{2+r})^{(1)}=R^{(2)}_{2+r}$.  \\
It remains to find out the CC for $(u,v)$ in the initial inhomogeneous system. As we have used two prolongations in order to exhibit $R^{(2)}_2$, we have second order formal derivatives of $u$ and $v$ in the right members. Now, as we have an involutive system, we have first order CC for the new right members and could hope therefore for third order generating CC. However, we have successively the $4$ CC:  \\
\[ \left\{ \begin{array}{l}
y_{233}=d_3(v_3-u_1)= d_2u \Rightarrow   \fbox {$  v_{33}-u_{13} -u_2=0 $}           \\
y_{223}= d_3(v_2-v_{13}+u_{11})= d_2(v_3-u_1)  \Rightarrow   \fbox {$ v_{133}  - u_{113} -u_{12}=0 $}  \\
y_{133}+y_{23}= d_3 v= d_1u +(v_3 - u_1) \Rightarrow \fbox {$ 0=0 $}  \\
y_{123}+y_{22}= d_2v = d_1(v_3-u_1) +(v_2 -v_{13}+u_{11}) \Rightarrow  \fbox {$ 0=0 $}       
\end{array} \right. \]
It follows that we have {\it only} one second and one third order CC:  \\
\[  v_{33} -u_{13} -u_2=0    , \hspace{2cm}   v_{133} -u_{113} - u_{12}=0  \]
but, {\it surprisingly}, we are left with the {\it only} generating second order CC $v_{33} -u_{13} -u_2=0$ which is coming from the fact that the operator $P$ commutes with the operator $Q$. \\

We let the reader prove as an exercise (See [14],[20] for details) that $dim(R_{r+2})=4r + 8, \forall r\geq 0$ and thus $ dim(R_3)=12, \,\, dim(R_4)= 16$ in the following commutative and exact diagrams where $E$ is the trivial vector bundle with $dim(E)=1$ and $dim(g_{r+2})=r+4, \forall r\geq 0$:  \\
  \[  \begin{array}{rcccccccl}
    &  0  &  &  0  &  &  0  &  &   &   \\
     &  \downarrow &  & \downarrow &  &  \downarrow  &  &   &     \\
0 \rightarrow  &  g_4  & \rightarrow & S_4T^*\otimes E & \rightarrow &  S_2T^*\otimes F_0 & \rightarrow &   h_2 &  \rightarrow 0 \\
&  \downarrow &  & \downarrow &  &  \downarrow  &  &  \downarrow    &     \\
0 \rightarrow &  R_4 & \rightarrow  & J_4(E) & \rightarrow  &  J_2(F_0) &  \rightarrow &  F_1  &  \rightarrow 0  \\
    &  \downarrow &  & \downarrow &  &  \downarrow  &  & \downarrow   &     \\
0 \rightarrow &  R_3 & \rightarrow  & J_3(E) & \rightarrow  &  J_1(F_0) &  \rightarrow &  0  &  \\ 
   &  &  & \downarrow &  &  \downarrow  &  &   & \\
   &  &  &  0  && 0  && &
\end{array}   \]
\[  \begin{array}{rcccccccl}
    &  0  &  &  0  &  &  0  &  &   &   \\
     &  \downarrow &  & \downarrow &  &  \downarrow  &  &   &     \\
0 \rightarrow  &  6  & \rightarrow &15 & \rightarrow & 12 & \rightarrow &  3&  \rightarrow 0 \\
&  \downarrow &  & \downarrow &  &  \downarrow  &  &  \downarrow    &     \\
0 \rightarrow &  16 & \rightarrow  & 35  & \rightarrow  & 20  &  \rightarrow &  1  &  \rightarrow 0  \\
    &  \downarrow &  & \downarrow &  &  \downarrow  &  & \downarrow   &     \\
0 \rightarrow & 12 & \rightarrow  &  20   & \rightarrow  & 8  &  \rightarrow &  0  &  \\ 
   &  &  & \downarrow &  &  \downarrow  &  &   & \\
   &  &  &  0  &  & 0  &  &  &
\end{array}   \]
We have thus the formally exact sequence:  \\
\[             0 \rightarrow  \Theta  \rightarrow  E \underset 2{\stackrel{{\cal{D}}}{\longrightarrow} }F_0 
\underset 2{\stackrel{{\cal{D}}_1}{\longrightarrow} } F_1 \rightarrow 0   \]
or, equivalently, the exact sequence of differential modules over $D$ as before:  \\
\[      0 \rightarrow  D  \underset 2 {\stackrel{}{\longrightarrow} } D^2 \underset 2 {\stackrel{Ê}{\longrightarrow } } D \stackrel{p}{ \longrightarrow} M   \rightarrow 0  \]
which is nevertheless {\it not} a Janet sequence because $R_2$ is {\it not} involutive.  \\

\noindent
{\bf MOTIVATING EXAMPLE 1.5}: We now prove that the case of variable coefficients can lead to strikingly different results, even if we choose them in the differential field $K=\mathbb{Q}(x^1,x^2,x^3)$ of rational functions in the coordinates that we shall meet in the study of the S and K metrics. In order to justify this comment, let us consider the simplest situation met with the second order system $R_2\subset J_2(E)$:  \\
\[ \fbox{ $ R_2 \subset J_2(E) \hspace{1cm}  \left\{ y_{33}=u, \,\,\,   y_{13} + x^3 y_2=v\hspace{1cm} \right. $}   \]
We may consider successively the following systems of decreasing dimensions $8>7>6>4$:  \\
\[ \fbox{ $  R'_2=R^{(1)}_2\subset R_2 \hspace{1cm}  \left\{ y_{33}=u,\,\,\,x^3y_{23} + y_2 = v_3 - u_1, \,\,\, y_{13} + x^3y_2= v  \right.\hspace{2cm} $}  \]
\[  \fbox{  $ R^{''}_2 =R^{(2)}_2 \subset R'_2 \hspace{1cm}
 \left\{ \begin{array}{l} 
 y_{33} = u, \,\,\, 2y_{23} =v_{33} - u_{13} - x^3u_2, \,\,\, y_{13} +x^3y_2 = v, \\
 2 y_2  =  -x^3v_{33} + x^3 u_{13} + (x^3)^2 u_2 + 2 v_3 - 2 u_1 
 \end{array} \right.   \hspace{1cm} $ }       \]
\[  \fbox{ $  R^{'''}_2 =R^{(3)}_2 \subset R^"_2 \hspace{1cm}
 \left\{ \begin{array}{l} 
 y_{33} = u, \,\,\, 2y_{23} =v_{33} - u_{13} - x^3u_2, \,\,\, y_{13} +x^3y_2 = v, \\
 2x^3y_{22}= - v_{133} + u_{113} + x^3 u_{12} + 2 v_2   \\
 2 y_{12}= - x^3v_{133} + x^3 u_{113} + (x^3)^2 u_{12} + 2 v_{13} - 2 u_{11}  \\
 2 y_2  =  -x^3v_{33} + x^3 u_{13} + (x^3)^2 u_2 + 2 v_3 - 2 u_1 
 \end{array} \right.  \hspace{1cm} $ }        \]
The last system is involutive with the following Janet tabular:  \\
\[  \left \{ \begin{array}{lcl}
  y_{33} &  =  & 0 \\
  y_{23}  &  =  & 0  \\
  y_{22}  &  =  &  0  \\
  y_{13}  &  =  &  0  \\
  y_{12} & = & 0  \\
  y_2      & = & 0
  \end{array} \right. \fbox{ $\begin{array}{lll}
  1 & 2 & 3  \\
  1 & 2 & \bullet  \\
  1 & 2 & \bullet   \\
  1  &  \bullet &  \bullet \\
  1  &  \bullet  &  \bullet  \\
  \bullet & \bullet & \bullet
  \end{array} $ }   \]
The generic solution is of the form $y=b(x^1) + c x^3$ and it is rather striking that such a system has constant coefficients (This will be {\it exactly} the case of the S and K metrics but similar examples can be found in [10]). We could hope for $9$ generating CC up to order $4$ but tedious computations, left to the reader as a tricky exercise, prove that we have in fact, as before, only $2$ generating {\it third order} CC described by the following involutive system, namely:  
\[    \fbox{ $ A \equiv v_{333} - u_{133} - x^3 u_{23} - 3 u_2 =0  $  }     \]
\[   \fbox{ $  B \equiv v_{133} - u_{113} - (x^3)^2 v_{233} + (x^3)^2 u_{123} + 2x^3 v_{23} +(x^3)^3 u_{22} - 3 x^3 u_{12} - 2 v_2 = 0   $ }  \]
satisfying the only first order CC:  
\[   \fbox{  $   d_3 B - d_1 A + (x^3)^2 d_2 A =0 $ }  \]
Introducing the ring $D=K[d_1,d_2,d_3]=K[d]$ of differential operators with coefficients in $K$, we obtain the sequence of $D$-modules:  \\
\[       0 \longrightarrow D \underset 1 {\stackrel{}{\longrightarrow}} D^2 \underset 3{\stackrel{ }{\longrightarrow}} D^2  \underset 2 {\stackrel{}{\longrightarrow}} D  \stackrel{p}{\longrightarrow} M \longrightarrow 0  \]
where the order of an operator is written under its arrow. This example proves that even a slight modification of the parameter can change the corresponding differential resolution.  \\

\noindent
{\bf MOTIVATING EXAMPLE 1.6}: We comment a tricky example first provided by M. Janet in $1920$, that we have studied with details in ([9],[11]). Using jet notations with $n=3, m=1, q=2, K=\mathbb{Q}(x^2)$, let us consider the inhomogeneous second order system:  \\
\[ R_2 \subset J_2(E) \hspace{2cm}  \left\{  y_{33} - x^2y_{11}= u, \,\,\,\,\, y_{22}=v \right. \hspace{4cm}\]
We let the reader prove that the space of solutions has dimension $12$ over $\mathbb{Q}$ and that we have $r=0, s=5$ in such a way that $R^{(5)}_2$ is involutive and even finite type with a zero symbol. Accordingly, we have $dim(R^{(5)}_2)=12$. Passing to the differential module point of view, it follows that $dim_K(M)=12$ and $rk_D(M)=0$. According to the general results presented, we have thus to use $5$ prolongations and could therefore wait for CC up to order ... $6$ !!!. In fact, {\it and we repeat that there is no hint at all for predicting this result in any intrinsic way}, we have only two generating CC, one of order $3$ and ... one of order $6$ indeed, namely:  \\
\[   \fbox{$ A \equiv   v_{233} - x^2 v_{112}- u_{222} - 3 v_{11} = 0 $} \]
\[   \fbox{$ \begin{array}{rcl}
 B&\equiv &  v_{333333} - x^2 v_{113333} - u_{223333} \\
  &  & - 2 x^2 (v_{113333} - x^2 v_{111133} - u_{112233}  )  \\
  &  &  + (x^2)^2 ( v_{111133}  - x^2 v_{111111} - u_{111122})  \\
    &  &  - 2 u_{11233}  + 2 x^2 u_{11112} - 2 u_{1111} = 0
    \end{array} $}  \]
satisfying the only fourth order CC:  \\
\[  \fbox{ $ C  \equiv A_{3333} - 2 x^2 A_{1133} + (x^2)^2 A_{1111} - B_2 =0 $ } \]
It follows that we have the unexpected differential resolution:  \\
\[      0  \rightarrow  D  \underset 4 {\stackrel{ }{\longrightarrow}} D^2  \underset 6 {\stackrel{ }{\longrightarrow}}D^2 \underset 2 {\stackrel{ }{\longrightarrow}} D  \stackrel{p}{\longrightarrow}  M  \rightarrow 0  \]
with Euler-Poincar\'{e} characteristic $rk_D(M)=1-2 +2 -1=0$ as expected. In addition, if we introduce a constant parameter $a$ by replacing the coefficient $x^2$ by $ax^2$, we obtain $2a y_{112}= v_{33} - ax^2 v_{11} - u_{22}$ and obtain the same conclusions as before. We point out the fact that, when $a=0$, the system $y_{33}=u, y_{22}=v$, which is trivially FI because it is homogeneous, has a symbol $g_2$ which is {\it neither} involutive (otherwise it should admit a first order CC), {\it nor even} $2$-acyclic because we have the parametric jets:  
\[par_2=(y_{11},y_{12},y_{13},y_{23}), \,\,\, par_3=(y_{111}, y_{112}, y_{113},y_{123}), \,\,\, par_4=(y_{1111},y_{1112},y_{1113}, y_{1123})\]
and the long $\delta$-sequence:  \\
\[  0  \rightarrow g_4 \stackrel{\delta}{\longrightarrow} T^*\otimes g_3 \stackrel{\delta}{\longrightarrow} {\wedge}^2T^*\otimes g_2 \stackrel{\delta}{\longrightarrow} {\wedge}^3T^*\otimes T^* \rightarrow 0  \]
\[  0  \rightarrow 4 \stackrel{\delta}{\longrightarrow}12\stackrel{\delta}{\longrightarrow} 12\stackrel{\delta}{\longrightarrow} 3 \rightarrow 0  \]
in which $dim(B^2(g_2))=12 - 4= 8, dim(Z^2(g_2))= 12-3=9 \Rightarrow dim(H^2(g_2)=9-8=1 \neq 0$.  \\
However, $g_3$ is involutive with the following Janet tabular for the vertical jets $(v_{ijk}) \in S_3T^*$:  \\
 \[  \left \{ \begin{array}{lcl}
  v_{333} &  =  & 0 \\
  v_{233}  &  =  & 0  \\
  v_{223}  &  =  &  0  \\
  v_{222}  &  =  &  0  \\
  v_{133} & = & 0  \\
  v_{122}     & = & 0
  \end{array} \right. \fbox{ $\begin{array}{lll}
  1 & 2 & 3  \\
  1 & 2 & \bullet  \\
  1 & 2 & \bullet   \\
  1  &   2  & \bullet  \\
  1  &  \bullet &  \bullet \\
  1  &  \bullet  &  \bullet  
  \end{array} $ }   \]
Accordingly, $R_3$ is thus involutive and the only CC $v_{33} - u_{22}=0$ is of order $2$ because we need one prolongation only to reach involution and thus $2$-acyclicity. \\

\noindent
{\bf MOTIVATING EXAMPLE 1.7}: With $m=1, n=2, q=2, K= \mathbb{Q}$, let us consider the inhomogeneous second order system:   \\
\[      y_{22}=u , \hspace{3cm}  y_{12}-y=v   \]
We obtain at once through crossed derivatives  $   y=u_{11}-v_{12}-v  $ and, by substituting, two fourth order CC for $(u,v)$, namely:\\
\[ \fbox { $ A \equiv  u_{1122}-v_{1222}-v_{22}-u = 0, \,\,\,\, B  \equiv  u_{1112}-u_{11}-v_{1122}  =  0   $ }\]
satisfying  \fbox {$B_{12}+B-A_{11}=0 $}. However,  we may also obtain a single CC for $(u,v)$, namely \fbox {$  C \equiv d_{12}u-u-d_{22}v=0 $} and we check at once $   A=d_{12}C+C, B=d_{11}C  $  while $ C=d_{22}B-d_{12}A+A $. We let the reader prove that $dim(R_{2+r})=4, \forall r\geq 0$. Hence, if $(A,B)$ is a section of $F_1$ while $C$ is a section of $F'_1$, the jet prolongation sequence: \\
\[  0 \rightarrow R_6  \rightarrow J_6(E) \rightarrow J_4(F_0) \rightarrow F_1 \rightarrow 0  \]
\[ 0 \rightarrow 4  \rightarrow 28 \rightarrow 30 \rightarrow 2 \rightarrow 0  \]
is {\it not} formally exact because $4 - 28 + 30 -2=4 \neq 0$, while the corresponding long sequence:  \\
\[ 0 \rightarrow R_{r+4}  \rightarrow J_{r+4}(E) \rightarrow J_{r+2}(F_0) \rightarrow J_r(F'_1) 
\rightarrow 0\]
\[ 0 \rightarrow 4  \rightarrow (r+5)(r+6)/2 \rightarrow (r+3)(r+4) \rightarrow (r+1)(r+2)/2 \rightarrow 0  \]
is indeed formally exact because  $4 - \frac{r^2+11r+30)}{2}+ (r^2+7r+12) - \frac{(r^2+3r+2)}{2}=0  $ but {\it not} strictly exact because $R_2$ is quite far from being FI as we have even $R^{(4)}_2=0$.  \\

It follows from these examples and the many others presented in ([20]) that we cannot agree with ([1-4]). Indeed, it is clear that one can use successive 
prolongations  in order to look for CC of order $1,2,3,...$ and so on, selecting each time the new generating ones  and knowing that Noetherian arguments will stop such a procedure ... after a while !. However, it is clear that, as long as the numbers $r$ and $s$ are not known, it is not {\it effectively} possible to decide in advance about the maximum order that must be reached. Therefore, it becomes clear that exactly the same procedure {\it must} be applied when looking for the CC of the various Killing operators we want to study, the problem becoming a " mathematical " one, {\it surely not} a " physical " one.  \\

\noindent
{\bf IMPORTANT REMARK 1.8}: Taking the adjoint operators, it is essential to notice that $ad(\cal{D})$ generates the CC of $ad({\cal{D}}_1)$ when $a\neq 0$, a result leading to $ext^1(M)=0$ but this is not true when $a=0$, a result leading to $ext^1(M)\neq 0$ ([10],[18],[21],[22]). Hence we discover on such an example that the intrinsic properties of a system with constant or even variable coefficients may {\it drastically} depend on these coefficients, even though the correspondig systems do not appear to be quite different at first sight. Accordingly, we have:  \\
WHEN A SYSTEM IS NOT FI, NOTHING CAN BE SAID " {\it A PRIORI} " ABOUT THE CC, THAT IS WITHOUT ANY EXPLICIT COMPUTATION OR COMPUTER ALGEBRA. \\

\noindent
Comparing the sequences obtained in the previous examples, we may state:  \\

\noindent
{\bf DEFINITION 1.9}: A differential sequence is said to be {\it formally exact} if it is exact on the jet level composition of the prolongations involved. A formally exact sequence is said to be {\it strictly exact} if all the operators/systems involved are FI (See [14] for more details). A strictly exact sequence is called {\it canonical} if all the operators/systems are involutive. The only known canonical sequences are the Janet and Spencer sequences that can be defined independently from each other.\\

With canonical projection ${\Phi}_0=\Phi:J_q(E) \Rightarrow J_q(E)/R_q=F_0$, the various prolongations are described by the following commutative and exact {\it introductory diagram}:  \\
\[  \begin{array}{rcccccccl}
    &  0  &  &  0  &  &  0  &  &   &   \\
     &  \downarrow &  & \downarrow &  &  \downarrow  &  &   &     \\
0 \rightarrow  &  g_{q+r+1}  & \rightarrow & S_{q+r+1}T^*\otimes E & \rightarrow &  S_{r+1}T^*\otimes F_0 & \rightarrow &   h_{r+1} &  \rightarrow 0 \\
&  \downarrow &  & \downarrow &  &  \downarrow  &  &  \downarrow    &     \\
0 \rightarrow &  R_{q+r+1} & \rightarrow  & J_{q+r+1}(E) & \stackrel{{\rho}_{r+1}(\Phi)}{\longrightarrow}  &  J_{r+1}(F_0) &  \rightarrow &  
Q_{r+1}  &  \rightarrow 0  \\
    &  \downarrow &  & \downarrow &  &  \downarrow  &  & \downarrow   &     \\
0 \rightarrow &  R_{q+r} & \rightarrow  & J_{q+r}(E) & \stackrel{{\rho}_r(\Phi)}{\longrightarrow}  &  J_r(F_0) &  \rightarrow &  Q_r  & 
\rightarrow 0 \\ 
   &  &  & \downarrow &  &  \downarrow  &  & \downarrow  & \\
   &  &  &  0  && 0  && 0  &
\end{array}   \]
Chasing along the diagonal of this diagram while applying the standard "{\it snake}" lemma, we obtain the useful {\it long exact connecting sequence}:  \\
 \[  0  \rightarrow  g_{q+1}  \rightarrow R_{q+1}  \rightarrow R_q  \rightarrow h_1 \rightarrow Q_1  \rightarrow 0  \]
which is thus connecting in a tricky way FI ({\it lower left}) with CC ({\it upper right}).  \\

 We finally recall the {\it Fundamental Diagram I} that we have presented in many books and papers, relating the (upper) {\it canonical Spencer sequence} to the (lower) {\it canonical Janet sequence}, that only depends on the left commutative square ${\cal{D}}=\Phi \circ j_q$ with $\Phi = {\Phi}_0$ when one has an involutive system $R_q\subseteq J_q(E)$ over $E$ with $dim(X)=n$ and $j_q:E \rightarrow J_q(E)$ is the derivative operator 
 up to order $q$:  \\

 \[ \footnotesize  \begin{array}{rcccccccccccccl}
 &&&&& 0 &&0&&0&&  &  &0&  \\
 &&&&& \downarrow && \downarrow && \downarrow & &  &   & \downarrow &  \\
  & 0& \rightarrow& \Theta &\stackrel{j_q}{\rightarrow}&C_0 &\stackrel{D_1}{\rightarrow}& C_1 &\stackrel{D_2}{\rightarrow} & C_2 &\stackrel{D_3}{\rightarrow}& ... &\stackrel{D_n}{\rightarrow}& C_n &\rightarrow 0 \\
  &&&&& \downarrow & & \downarrow & & \downarrow & &  &&\downarrow &     \\
   & 0 & \rightarrow & E & \stackrel{j_q}{\rightarrow} & C_0(E) & \stackrel{D_1}{\rightarrow} & C_1(E) &\stackrel{D_2}{\rightarrow} & C_2(E) &\stackrel{D_3}{\rightarrow} & ... &\stackrel{D_n}{\rightarrow} & C_n(E) &   \rightarrow 0 \\
   & & & \parallel && \hspace{6mm}\downarrow  {\Phi}_0& & \hspace{6mm}\downarrow  {\Phi}_1& & \hspace{6mm}\downarrow {\Phi}_2 & &  & & \hspace{6mm} \downarrow  {\Phi}_n & \\
   0 \rightarrow & \Theta &\rightarrow & E & \stackrel{\cal{D}}{\rightarrow} & F_0 & \stackrel{{\cal{D}}_1}{\rightarrow} & F_1 & \stackrel{{\cal{D}}_2}{\rightarrow} & F_2 & \stackrel{{\cal{D}}_3}{\rightarrow} & ... &\stackrel{{\cal{D}}_n}{\rightarrow} & F_n & \rightarrow  0 \\
   &&&&& \downarrow & & \downarrow & & \downarrow & &  &  &\downarrow &   \\
   &&&&& 0 && 0 && 0 && &&0 &  
   \end{array}     \]
We shall use this result, first found exactly $40$ years ago ([7]) but never acknowledged, in order to provide a critical study of the comparison between the S and K metrics.  \\

\noindent
{\bf EXAMPLE 1.10}: The Janet tabular in Example 1.4 with $a=1$ provides the fiber dimensions: \\
\[ \begin{array}{rccccccccccccl}
 &&&&& 0 &&0&&0&& 0 &  \\
 &&&&& \downarrow && \downarrow && \downarrow &   & \downarrow &  \\
  & 0& \rightarrow& \Theta &\stackrel{j_2}{\rightarrow}& 6 &\stackrel{D_1}{\rightarrow}& 16 &\stackrel{D_2}{\rightarrow} & 14&\stackrel{D_3}{\rightarrow}& 4 &\rightarrow 0 \\
  &&&&& \downarrow & & \downarrow & & \downarrow & &\parallel &     \\
   & 0 & \rightarrow & 1 & \stackrel{j_2}{\rightarrow} & 10 & \stackrel{D_1}{\rightarrow} & 20 &\stackrel{D_2}{\rightarrow} &15 &\stackrel{D_3}{\rightarrow}  & 4 &   \rightarrow 0 \\
   & & & \parallel && \hspace{6mm}\downarrow  {\Phi}_0& & \hspace{6mm}\downarrow  {\Phi}_1& & \hspace{6mm}\downarrow {\Phi}_2 & &  \downarrow  & \\
   0 \rightarrow & \Theta &\rightarrow & 1 & \stackrel{\cal{D}}{\rightarrow} & 4 & \stackrel{{\cal{D}}_1}{\rightarrow} & 4 & \stackrel{{\cal{D}}_2}{\rightarrow} & 1 & \rightarrow  & 0  & \\
   &&&&& \downarrow & & \downarrow & & \downarrow & & &   \\
   &&&&& 0 && 0 && 0& &  
   \end{array}     \]
We notice that $6 - 16 +14 -4=0$, $1 - 10 + 20 - 15 + 4=0$ and $1-4 +4 -1=0$. In this diagram, the Janet sequence seems simpler than the Spencer sequence but, sometimes as we shall see, it is the contrary and there is no rule. We invite the reader to treat similarly the cases $a=0$ and $a=x^3$. \\

\noindent 
{\bf 2) SCHWARZSCHILD VERSUS KERR} \\

\noindent
{\bf  a) SCHWARZSCHILD METRIC}  \\

In the Boyer-Lindquist (BL) coordinates $(t,r,\theta, \phi)=(x^0,x^1,x^2,x^3)$, the Schwarzschild metric is $\omega= A(r)dt^2-(1/A(r))dr^2 - r^2d{\theta}^2 -r^2{sin}^2(\theta)d{\phi}^2$ and $\xi={\xi}^id_i \in T$, let us  introduce ${\xi}_i={\omega}_{ri}{\xi}^r$ with the $4$ {\it formal derivatives} $(d_0=d_t, d_1=d_r,d_2= d_{\theta}, d_3=d_{\phi})$. With speed of light $c=1$ and $A=1-\frac{m}{r}$ where $m$ is a constant, the metric can be written in the diagonal form:  \\
\[ \left(    \begin{array}{cccc}
A & 0  &  0  &  0   \\
0 & -1/A & 0 & 0 \\
0 & 0 & -r^2 & 0 \\
0 & 0 & 0 & -r^2sin^2(\theta)
\end{array}  \right)   \]
with a surprisingly simple determinant $det(\omega)= - r^4sin^2(\theta)$.  \\
Using the notations of differential modules or jet theory, we may consider the infinitesimal Killing equations:   \\
\[ \Omega \equiv {\cal{L}}(\xi)\omega=0 \hspace{2mm} \Leftrightarrow \hspace{2mm} {\Omega}_{ij}\equiv d_i{\xi}_j +d_j{\xi}_i  - 2 {\gamma}^r_{ij}{\xi}_r=0  \hspace{2mm} \Leftrightarrow \hspace{2mm} {\Omega}_{ij}\equiv {\omega}_{rj}{\xi}^r_i + {\omega}_{ir}{\xi}^r_j + {\xi}^r {\partial}_r{\omega}_{ij} =0 \]
where we have introduced the Christoffel symbols $\gamma$ through he standard Levi-Civita isomorphism $j_1(\omega)\simeq (\omega, \gamma)$ while  setting $A'={\partial}_rA$ in the differential field $K$ of coefficients ([19]). As in the Macaulay example just considered and in order to avoid any further confusion between sections and derivatives, we shall use the sectional point of view and rewrite the previous \fbox{$10$} equations in the symbolic form $\Omega \equiv  L({\xi}_1)\omega \in S_2T^* $ where $L$ is the {\it formal Lie derivative}:  \\ 
\[ R_1\subset J_1(T) \,\,\,  \left\{ \begin{array}{rcrl}
{\Omega}_{33} &  \equiv  &  - 2 r^2 {sin}^2(\theta) & \fbox{${\xi}^3_3$} - 2r sin^2(\theta){\xi}^1 - 2r^2sin(\theta)cos(\theta) {\xi}^2 =0  \\
{\Omega}_{23  } &  \equiv  & -r^2 & \fbox{${\xi}^2_3$} - r^2 {sin}^2(\theta) {\xi}^3_2   =0   \\
{\Omega}_{13  } &  \equiv  &  - \frac{1}{A}& \fbox{$ {\xi}^1_3$}  - r^2 {sin}^2(\theta) {\xi}^3_1  =0  \\
{\Omega}_{03  } &  \equiv  &  A & \fbox{${\xi}^0_3$} - r^2 {\sin}^2(\theta) {\xi}^3_0    =0   \\
{\Omega}_{22} &  \equiv  & - 2r^2 & \fbox{${\xi}^2_2$}  - 2r {\xi}^1 =0  \\
{\Omega}_{12  } &  \equiv  &  - \frac{1}{A}& \fbox{$ {\xi}^1_2$} - r^2{\xi}^2_1  =0   \\
{\Omega}_{02  } &  \equiv  &  A & \fbox{$ {\xi}^0_2 $} - r^2 {\xi}^2_0     =0  \\
{\Omega}_{11}  &  \equiv  &  - \frac{2}{A}& \fbox{${\xi}^1_1$} + \frac{A'}{A^2}{\xi}^1=0   \\
{\Omega}_{01}  &  \equiv  &  - \frac{1}{A}& \fbox{${\xi}^1_0$} + A{\xi}^0_1 =0  \\
{\Omega}_{00} &  \equiv  &   2A&  \fbox{${\xi}^0_0 $}  + A'{\xi}^1=0
\end{array} \right.  \]

Though this system $R_1\subset J_1(T)$ has 4 equations of class $3$, 3 equations of class $2$, 2 equations of class $1$ and 1 equation of class $0$, it is far from being involutive because it is finite type with second symbol $g_2=0$ defined by the 40 equations $v^k_{ij}=0$ in the initial coordinates. From the symetry, it is clear that such a system has {\it at least} 4 solutions, namely the time translation ${\partial}_t \leftrightarrow {\xi}^0=1 \Leftrightarrow {\xi}_0= A$ and, using cartesian coordinates $(t,x,y,z)$, the 3 space rotations $y{\partial}_z-z{\partial}_y, z{\partial}_x - x{\partial}_z, x{\partial}_y - y{\partial}_x$. \\
We obtain in particular, modulo $\Omega$:  \\
\[  {\xi}^0_0= - \frac{A'}{2A}{\xi}^1, {\xi}^1_1= + \frac{A'}{2A} {\xi}^1, {\xi}^2_2= - \frac{1}{r}{\xi}^1, {\xi}^3_3= - \frac{1}{r}{\xi}^1 - cot(\theta){\xi}^2  \,\, \Rightarrow \,\,  {\xi}^0_0 + {\xi}^1_1=0, {\xi}^2_2 + {\xi}^3_3= - cot(\theta) {\xi}^2 \]

We may also write the Schwarzschild metric in cartesian coordinates as:ÊÊ\\
\[   \omega=A(r)dt^2 + (1-\frac{1}{A(r)})dr^2 - (dx^2+dy^2+dz^2), \hspace{1cm}  rdr=xdx+ydy+zdz  \]
and notice that the $3\times 3$ matrix of components of the three rotations has rank equal to $2$, a result leading surely, before doing any computation, to the existence of {\it one and only one} zero order Killing equation $r{\xi}^r=x{\xi}^x+y{\xi}^y+z{\xi}^z=0 \Rightarrow {\xi}^1={\xi}^r=0$.  Such a result also amounts to say that the spatial projection of any Killing vector on the radial spatial unit vector $(x/r,y/r,z/r)$ vanishes beause $r$ must stay invariant. \\

However, {\it as we are dealing with sections}, ${\xi}^1=0$ implies ${\xi}^0_0=0,{\xi}^1_1=0,{\xi}^2_2=0$ ...  but {\it NOT} ({\it care}) ${\xi}^1_0=0$, these later condition being only brought by one additional prolongation and we have the strict inclusions  $R^{(3)}_1\subset R^{(2)}_1 \subset R^{(1)}_1=R_1$ that we rename as $R^"_1 \subset R'_1 \subset R_1$. Hence, it remains to determine the dimensions of these subsystems and their symbols, exactly like in the Macaulay example. We shall prove in the next section that two prolongations bring the five new equations:  \\
\[  \fbox{  ${\xi}^1=0, \,\,\, {\xi}^1_2=0, \,\,\, {\xi}^1_3=0, \,\,\, {\xi}^0_2=0, \,\,\, {\xi}^0_3=0 $ }  \]
and a new prolongation {\it only} brings the single equation $\fbox { $ {\xi}^1_0=0$}$.

Knowing that $dim(R_1)=dim(R_2)=10$, $ dim(R_3) = 5$, $dim(R_4)= 4$, we have thus obtained the \fbox{$15$} equations defining $R'_1=R^{(2)}_1$ with $dim(R'_1)=20-15=5$ and let the reader draw the corresponding Janet tabular for the $4$ equations of class $3$, the $4$ equations of class $1$, the $3$ equations of class $0$ and the $3$ equations of class $2$. The symbol $g'_1$ has the two parametric jets $(v^3_2,v^1_0)$ and is not $2$-acyclic. Adding ${\xi}^1_0=0 \Leftrightarrow {\xi}^0_1=0$, we finally achieve the PP procedure with the \fbox{$16$} equations defining the system $R^"_1=R^{(3)}_1$ with $dim(R^"_1)=20-16=4$, namely:  \\
\[ R^"_1\subset R'_1 \subset R_1 \subset J_1(T) \left\{ \begin{array}{l}
{\xi}^3_3 + cot(\theta) {\xi}^2 =0  \\
{\xi}^2_3 + sin^2(\theta){\xi}^3_2=0  \\
{\xi}^1_3  =0  \\
{\xi}^0_3 =0  \\
{\xi}^3_1 =0  \\
{\xi}^2_1  =0  \\
{\xi}^1_1 =0  \\
{\xi}^0_1  =0 \\
{\xi}^3_0 =0  \\
{\xi}^2_0 =0 \\
{\xi}^1_0 =0  \\
{\xi}^0_0 =0  \\
{\xi}^2_2 =0 \\
{\xi}^1_2 =0  \\
{\xi}^0_2  =0  \\
{\xi}^1=0
\end{array} \right.  \fbox{ $ \begin{array}{cccc}
2& 0 & 1 & 3  \\
2 & 0 &1 & 3  \\
2 & 0 & 1 & 3 \\
2 & 0 & 1 & 3 \\
2 & 0 & 1 & \bullet  \\
2 & 0 & 1 & \bullet  \\
2& 0 & 1 & \bullet  \\
2 & 0 & 1 & \bullet  \\
2 & 0 & \bullet & \bullet  \\
2 & 0  & \bullet & \bullet  \\
2 & 0 & \bullet & \bullet   \\
2  &  0  & \bullet & \bullet  \\
2 & \bullet & \bullet  & \times  \\
2 & \bullet & \bullet & \bullet  \\
2 & \bullet & \bullet & \bullet  \\
\bullet & \bullet & \bullet & \bullet
\end{array}  $ }   \]
and we have replaced by "$\times$" the only " {\it dot} " (non-multiplicative variable) that cannot provide vanishing crossed derivatives and thus involution of the symbol $g^"_1$ with the only parametric jets $(v^3_2, v^1_0)$.  It is easy to check that $R^"_1$, having minimum dimension equal to $4$, is formally integrable, though not involutive as it is finite type with $dim(g^"_1) = 16 - 15=1 \Rightarrow g^"_1\neq 0$ with parametric jet $v^3_2$ and to exhibit $4$ solutions linearly independent over the constants. We let the reader prove as an exercise that the dimension of the Spencer $\delta$-cohomology at ${\wedge}^2T^*\otimes g^"_1$ is $dim((H^2(g^"_1))=3\neq 0$ but we have proved in ([19]) that its restriction to $(x^2,x^3)$ is of dimension $1$ only. We obtain:  \\

\hspace*{2cm} THIS SYSTEM DOES NOT DEPEND ON $m$ ANY LONGER !.  \\

Denoting by $R^"_2\subset R_2\subset J_2(T)$ with $dim(R^"_2)=4$ the prolongation of $R'_1 \subset J_1(T)$, it is the involutive system provided by the {\it prolongation/projection} (PP) procedure, we are in position to construct the corresponding canonical/involutive (lower) Janet and (upper) Spencer sequences along the following {\it fundamental diagram I} that we recalled in the Introduction. In the present situation, the Spencer sequence is isomorphic to the tensor product of the Poincar\'{e} sequence by the underlying $4$-dimensional Lie algebra ${\cal{G}}$, namely:  \\
\[   {\wedge}^0T^*\otimes {\cal{G}} \stackrel{d}{\longrightarrow } {\wedge}^1T^*\otimes {\cal{G}} \stackrel{d}{\longrightarrow} ... \stackrel{d}{\longrightarrow} {\wedge}^4T^*\otimes {\cal{G}} \longrightarrow 0  \]
In this diagram, {\it not depending any longer on} $m$, we have now $C_r={\wedge}^rT^*\otimes R^"_2$ and ${\cal{D}}$ is of order $2$ like $j_2$ while {\it all} the other operators are of order $1$: \\
 \[   \begin{array}{rcccccccccccccl}
 &&&&& 0 &&0&&0&& 0 &  &0&  \\
 &&&&& \downarrow && \downarrow && \downarrow & & \downarrow &   & \downarrow &  \\
  & 0& \rightarrow& \Theta &\stackrel{j_2}{\rightarrow}&4 &\stackrel{D_1}{\rightarrow}& 16 &\stackrel{D_2}{\rightarrow} & 24&\stackrel{D_3}{\rightarrow}&16 &\stackrel{D_4}{\rightarrow}& 4 &\rightarrow 0 \\
  &&&&& \downarrow & & \downarrow & & \downarrow & & \downarrow &&\downarrow &     \\
   & 0 & \rightarrow & 4 & \stackrel{j_2}{\rightarrow} & 60 & \stackrel{D_1}{\rightarrow} & 160 &\stackrel{D_2}{\rightarrow} & 180 &\stackrel{D_3}{\rightarrow} & 96 &\stackrel{D_4}{\rightarrow} &20&   \rightarrow 0 \\
   & & & \parallel && \downarrow  & & \downarrow  & & \downarrow  & & \downarrow & & \downarrow  & \\
   0 \rightarrow & \Theta &\rightarrow & 4 & \underset 2{\stackrel{\cal{D}}{\rightarrow}} & 56 & \stackrel{{\cal{D}}_1}{\rightarrow} & 144 & \stackrel{{\cal{D}}_2}{\rightarrow} & 156 & \stackrel{{\cal{D}}_3}{\rightarrow} & 80 &\stackrel{{\cal{D}}_4}{\rightarrow} & 16 & \rightarrow  0 \\
   &&&&& \downarrow & & \downarrow & & \downarrow & & \downarrow &  &\downarrow &   \\
   &&&&& 0 && 0 && 0 &&0 &&0 &  
   \end{array}     \]
 We notice the vanishing of the Euler-Poincar\'{e} characteristics:  \\  
\[   4 - 16 +24 - 16 + 4=0, \hspace{2mm}  4 - 60 +160 - 180 + 96 - 20 =0, \hspace{2mm} 4 - 56 + 144 - 156 + 80 - 16 = 0   \]
We point out that, whatever is the sequence used or the way to describe ${\cal{D}}_1$, then $ad({\cal{D}}_1)$ is parametrizing the {\it Cauchy} operator $ad({\cal{D}})$ for the S metric. However, such an approach does not tell us {\it explicitly} what are the second and third order CC involved in the initial 
situation. \\

In actual practice, {\it all the preceding computations have been finally used to reduce the Poincar\'e group to its subgroup made with only one time translation and three space rotations} !. On the contrary, we have proved during almost fourty years that one {\it must} increase the Poincar\'e group ($10$ parameters), first to the Weyl group ($11$ parameters by adding $1$ dilatation) and finally to the conformal group of space-time ($15$ parameters by adding $4$ elations) while only dealing with he Spencer sequence in order to increase the dimensions of the Spencer bundles, thus the number $dim(C_0)$ of  {\it potentials} and the number $dim(C_1)$ of {\it fields} (Compare to [6]).  \\

\noindent
{\bf b) KERR METRIC}  \\

We now write the Kerr metric in Boyer-Lindquist coordinates:  \\
\[  \begin{array}{rcl}
ds^2  &  =  & \frac{{\rho}^2 - mr}{{\rho}^2}dt^2 - \frac{{\rho}^2}{\Delta} dr^2  -  {\rho}^2 d{\theta}^2  \\
  &    &  - \frac{2a m r sin^2(\theta)}{{\rho}^2} dtd\phi - (r^2+a^2 + \frac{mr a^2 sin^2(\theta)}{{\rho}^2})sin^2(\theta) d{\phi}^2          
  \end{array}   \]
where we have set $  \Delta= r^2  -mr +a^2 , \,\,  {\rho}^2=r^2 + a^2 cos^2(\theta) $ as usual and we check that:  \\
\[  a=0  \Rightarrow  ds^2= (1- \frac{m}{r})dt^2 - \frac{1}{1- \frac{m}{r}}dr^2 -r^2d{\theta}^2 - r^2 sin(\theta)^2 d{\phi}^2 \]
as a well known way to recover the Schwarschild metric. We notice that $t$ or $\phi$ do not appear in the coefficients of the metric and thus, as the maximum subgroup of invariance of the Kerr metric {\it must} be contained in the maximum subgroup of invariance of the Schwarzschild metric because of the above limit when $a\rightarrow 0$, we obtain the only possible $2$ infinitesimal generators $\{ {\partial}_t, {\partial}_{\phi}\}$ and we have the fundamental diagram I with fiber dimensions:  \\
\[   \begin{array}{rcccccccccccccl}
 &&&&& 0 &&0&&0&& 0 &  &0&  \\
 &&&&& \downarrow && \downarrow && \downarrow & & \downarrow &   & \downarrow &  \\
  & 0& \rightarrow& \Theta &\stackrel{j_2}{\rightarrow}&2 &\stackrel{D_1}{\rightarrow}& 8 &\stackrel{D_2}{\rightarrow} & 12 &\stackrel{D_3}{\rightarrow}& 8 &\stackrel{D_4}{\rightarrow}& 2 &\rightarrow 0 \\
  &&&&& \downarrow & & \downarrow & & \downarrow & & \downarrow &&\downarrow &     \\
   & 0 & \rightarrow & 4 & \stackrel{j_2}{\rightarrow} & 60 & \stackrel{D_1}{\rightarrow} & 160 &\stackrel{D_2}{\rightarrow} & 180 &\stackrel{D_3}{\rightarrow} & 96 &\stackrel{D_4}{\rightarrow} &20&   \rightarrow 0 \\
   & & & \parallel && \downarrow  & & \downarrow  & & \downarrow  & & \downarrow & & \downarrow  & \\
   0 \rightarrow & \Theta &\rightarrow & 4 & \stackrel{\cal{D}}{\rightarrow} & 58 & \stackrel{{\cal{D}}_1}{\rightarrow} & 152 & \stackrel{{\cal{D}}_2}{\rightarrow} & 168 & \stackrel{{\cal{D}}_3}{\rightarrow} & 88 &\stackrel{{\cal{D}}_4}{\rightarrow} & 18 & \rightarrow  0 \\
   &&&&& \downarrow & & \downarrow & & \downarrow & & \downarrow &  &\downarrow &   \\
   &&&&& 0 && 0 && 0 &&0 &&0 &  
   \end{array}     \]
with Euler-Poincar\'{e} characteristic $4 - 58 +152 - 168 + 88 - 18 = 0  $. Comparing the {\it surprisingly high} dimensions of the Janet bundles with the {\it surprisingly low} dimensions of the Spencer bundles needs no comment on the physical usefulness of the Janet sequence, despite its purely mathematical importance. In addition, using the same notations as in the preceding section, we shall prove that we have now the additional {\it zero order} equations ${\xi}^r=0, {\xi}^{\theta}=0$ produced by the non-zero components of the Weyl tensor and thus, {\it at best}, $dim(R^{(3)}_0)=2 \Leftrightarrow  dim((R^{(2)}_1)=2$ as these zero order equations will be obtained after {\it only} two prolongations. They depend on $j_2(\Omega)$ and we should obtain therefore eventually $dim(Q_2)=10+dim(R_3)\geq 12$ CC of order $2$ without any way way to know abut the desired third order CC.  \\ 

Using now cartesian space coordinates $(x,y,z)$ with ${\xi}^z=0, \,\,x {\xi}^x +y{\xi}^y=0$, we have only to study the following first order involutive system for ${\xi}^x=\xi$ with coefficients no longer depending on $(a,m)$, providing the only generator $x{\partial}_y - y {\partial}_x$: \\
\[ \left\{   \begin{array}{lcl}
{\Phi}^3 & \equiv & {\xi}_z =0 \\
{\Phi}^2 & \equiv &  {\xi}_y - \frac{1}{y}{\xi}  =0  \\
{\Phi}^1 & \equiv & {\xi}_x =0 
\end{array}  \right.    \fbox{  $  \begin{array} {ccc}
1 & 2 & 3   \\
1 & 2 & \bullet \\
1 & \bullet &  \bullet   
\end{array}  $  }  \]
and the fundamental diagram\  \\

\[   \begin{array}{rcccccccccccl}
 &&&&& 0 &&0&&0&  &0&  \\
 &&&&& \downarrow && \downarrow && \downarrow &    & \downarrow &  \\
  & 0& \rightarrow& \Theta &\stackrel{j_1}{\rightarrow}&1 &\stackrel{D_1}{\rightarrow}& 3 &\stackrel{D_2}{\rightarrow} & 3 &\stackrel{D_3}{\rightarrow}& 1 &\rightarrow 0 \\
  &&&&& \downarrow & & \downarrow & & \downarrow & &\parallel &     \\
   & 0 & \rightarrow & 1 & \stackrel{j_1}{\rightarrow} & 4 & \stackrel{D_1}{\rightarrow} & 6 &\stackrel{D_2}{\rightarrow} & 4 &\stackrel{D_3}{\rightarrow} & 1 &   \rightarrow 0 \\
   & & & \parallel && \downarrow  & & \downarrow  & & \downarrow  &  & \downarrow  & \\
   0 \rightarrow & \Theta &\rightarrow & 1 & \stackrel{\cal{D}}{\rightarrow} & 3 & \stackrel{{\cal{D}}_1}{\rightarrow} & 3 & \stackrel{{\cal{D}}_2}{\rightarrow} & 1 & \rightarrow & 0 & \\
   &&&&& \downarrow & & \downarrow & & \downarrow &  &   &   \\
   &&&&& 0 && 0 && 0 && &  
   \end{array}     \]
   
The involutive system produced by the PP procedure does not deend on $(m,a)$ any longer. Accordingly, this final result definitively proves that, {\it as far as differential sequences are concerned}: \\

\hspace*{15mm}THE ONLY IMPORTANT OBJECT IS THE GROUP, NOT THE METRIC   \\

\noindent
{\bf 3) SCHWARZSCHILD METRIC REVISITED}  \\

Let us now introduce the Riemann tensor $({\rho}^k_{l,ij})  \in {\wedge}^2T^*\otimes T^*\otimes T$ and use the metric in order to raise or lower the indices in order to obtain the purely covariant tensor $({\rho}_{kl,ij})\in {\wedge}^2T^*\otimes T^*\otimes T^*$. Then, using $r$ as an implicit summation index, we may consider the formal Lie derivative on sections:  \\
\[  R_{kl,ij}\equiv {\rho}_{rl,ij}{\xi}^r_k +{\rho}_{kr,ij}{\xi}^r_l+{\rho}_{kl,rj}{\xi}^r_i + {\rho}_{kl,ir}{\xi}^r_j + {\xi}^r {\partial}_r {\rho}_{kl,ij}=0  \]
that can be considered as an infinitesimal variation. As for the Ricci tensor $({\rho}_{ij}) \in S_2T^*$, we notice that ${\rho}_{ij}={\rho}^r_{i,rj}=0 \Rightarrow R_{ij}\equiv {\rho}_{rj}{\xi}^r_i + {\rho}_{ir}{\xi}^r_j +{\xi}^r{\partial}_r{\rho}_{ij}=0  $ though we have only: \\
\[ {\omega}^{rs}R_{ri,sj}=R_{ij} + {\omega}^{rs}{\rho}^t_{i,sj}{\Omega}_{st}\Rightarrow R_{ij}= R^r_{i,rj}= {\omega}^{rs} R_{ri,sj} \,\,\, mod(\Omega)\]
The $6$ non-zero components of the Riemann tensor are known to be:  \\
\[ \fbox{ $  \begin{array}{lll}
  {\rho}_{01,01}= + \frac{m}{r^3},& \,\,{\rho}_{02,02}= - \frac{m\,A}{2r}, &\,\,{\rho}_{03,03}= - \frac{m\,A\,sin^2(\theta)}{2r}  \\
  {\rho}_{12,12}= + \frac{m}{2r\,A},& \,\,{\rho}_{13,13}= + \frac{m\,sin^2(\theta)}{2r\,A},& \,\, {\rho}_{23,23}= - m\, r \, sin^2(\theta)  
\end{array}  $ } \]
First of all, we notice that:   \\
\[  {\xi}^0_0 + \frac{A'}{2A}{\xi}^1=0, \,\,  {\xi}^1_1 - \frac{A'}{2A}{\xi}^1=0 \,\, \Rightarrow \,\, {\xi}^0_0 + {\xi}^1_1=0 \]
\[   {\Omega}_{12} \equiv  - \frac{1}{A}{\xi}^1_2 - r^2 {\xi}^2_1=0, \,\,{\xi}^2_2 + \frac{1}{r}{\xi}^1=0  \]
We obtain therefore:  \\
\[  R_{01,01}\equiv 2{\rho}_{01,01} ({\xi}^0_0 + {\xi}^1_1) + {\xi}^r{\partial}_r({\rho}_{01,01})= {\xi}^1 {\partial }_1 {\rho}_{01,01}= - \frac{3m}{r^4}{\xi}^1= 0 \Rightarrow \fbox{${\xi}^1=0$} \]
\[  R_{02,02}\equiv 2 {\rho}_{02,02}({\xi}^0_0 + {\xi}^2_2) + {\xi}^r{\partial}_r({\rho}_{02,02}) \equiv (- \frac{mA}{r}(-\frac{A'}{2A}-\frac{1}{r}) -(\frac{mA}{2r})'){\xi}^1 = \frac{3mA}{2r^2}{\xi}^1=0  \]
Similarly, we also get:  \\
\[  R_{01,02} \equiv {\rho}_{01,01}{\xi}^1_2 + {\rho}_{02,02}{\xi}^2_1 +{\xi}^r{\partial}_r{\rho}_{01,02}=0 \,\,\Rightarrow \,\,  \frac{m}{r^3}{\xi}^1_2 - \frac{mA}{2r}{\xi}^2_1=0 \,\,  \Rightarrow \,\, \fbox{ ${\xi}^1_2=0 $}\]
\[  R_{01,03} \equiv {\rho}_{01,01}{\xi}^1_3 + {\rho}_{03,03}{\xi}^3_1 +{\xi}^r{\partial}_r{\rho}_{01,03}=0 \,\,\Rightarrow \,\,  \frac{m}{r^3}{\xi}^1_3 - \frac{mA{sin}^2(\theta)}{2r}{\xi}^3_1=0 \,\,  \Rightarrow \,\, \fbox{ ${\xi}^1_3=0 $}\]
and so on. We obtain for example, among the second order CC:  \\
\[   R_{01,01}\equiv - \frac{3m}{r^4}{\xi}^1=0, \,\,\, R_{02,02}\equiv \frac{3mA}{2r^2} {\xi}^1=0  \,\, \Rightarrow \,\,  R_{02,02} - \frac{r^2A}{2}R_{01,01}=0 \]
and thus, among the first prolongations, the third order CC that cannot be obtained by prolongation of the various second order CC while taking into account the Bianchi identities ([MSK]). Using the Spencer operator and the fact that ${\xi}^1 \in j_2(\Omega)$, we obtain indeed:  \\
\[ \fbox{ $ d_1{\xi}^1 - {\xi}^1_1=d_1{\xi}^1 - \frac{A'}{2A}{\xi}^1= 0, \,\,\, d_2{\xi}^1 - {\xi}^1_2=0, \,\,\, d_3{\xi}^1 - {\xi}^1_3=0 $ } \]

In addition, introducing ${\xi}^1\in j_2(\Omega)$ in the right member as in the motivating examples, we have $3$ PD equations for $({\xi}^2,{\xi}^3)$, namely: \\
\[  {\xi}^3_3 + cot(\theta){\xi}^2 = - \frac{1}{r}{\xi}^1 , \,\,\,  {\xi}^2_3 +{sin}^2(\theta){\xi}^3_2 =0, \,\,\, {\xi}^2_2 = - \frac{1}{r}{\xi}^1   \]
Using two prolongations and eliminating the third order jets, we obtain successively:  \\

\[  {\xi}^2_{233} + {sin}^2(\theta) {\xi}^3_{223}+ 2sin(\theta)cos(\theta) {\xi}^3_{23} =0   \]
\[  - {\xi}^2_{233}  = \frac{1}{r}{\xi}^1_{33}    \]
\[ -{sin}^2(\theta) {\xi}^3_{223} - sin(\theta)cos(\theta){\xi}^2_{22} + 2{\xi}^2_2 - 2 cot(\theta){\xi}^2 = - \frac{{\sin}^2(\theta)}{r}{\xi}^1_{22}   \]
\[   - 2sin(\theta)cos(\theta) {\xi}^3_{23}  - 2{cos}^2(\theta){\xi}^2_2 + 2 cot(\theta){\xi}^2=   \frac{2sin(\theta)cos(\theta)}{r}{\xi}^1_2  \]
\[   sin(\theta)cos(\theta) {\xi}^2_{22} = - \frac{\sin(\theta)cos(\theta)}{r}{\xi}^1_2   \]
\[   - 2 {sin}^2(\theta) {\xi}^2_2 = \frac{2{sin}^2(\theta)}{r} {\xi}^1   \]
Summing, we see that all terms in ${\xi}^2$ and ${\xi}^3$ disappear and that we are only left with terms in ${\xi}^1$, including in particular the second order jets ${\xi}^1_{22} , {\xi}^1_{33}$, namely:   \\
\[     {\xi}^1_{33} - {sin}^2(\theta) {\xi}^1_{22} - sin(\theta)cos(\theta) {\xi}^1_2 + 2 {sin}^2(\theta) {\xi}^1 =0   \]
Setting  $U={\xi}^1, V_2={\xi}^1_2, V_3={\xi}^1_3, W_2= {\xi}^0_2,W_3={\xi}^0_3 $ with $(U,V_2,V_3,W_2,W_3) \in j_2(\Omega)$, we obtain the 
new {\it strikingly unusual} third order CC for $\Omega$:   \\ 
\[ \fbox { $   d_3V_3 - {sin}^2(\theta)d_2V_2 - sin(\theta)cos(\theta) V_2 + 2 {sin}^2(\theta) U=0  $ } \]
However, in our opinion at least, we do not believe that such a purely "technical " relation could have any "physical " usefulness and let the reader compare it with the CC already found in ([19], Lemma 3.B.3). In addition and contrary to this situation, we have successively:   \\

\[ R_{01,12} \equiv  {\rho}_{01,10} {\xi}^0_2 + {\rho}_{21,12} {\xi}^2_0 + \xi \partial {\rho}_{01,12}= - \frac{m}{r^3}{\xi}^0_2 - \frac{m}{2rA}{\xi}^2_0=- \frac{3m}{2r^3} {\xi}^0_2=0 \Rightarrow \fbox{ ${\xi}^0_2=0$} \]           
\[ R_{01,13} \equiv  {\rho}_{01,10} {\xi}^0_3 + {\rho}_{31,13} {\xi}^3_0 + \xi \partial {\rho}_{01,13}= - \frac{m}{r^3}{\xi}^0_3 - \frac{m{sin}^2(\theta)}{2rA}{\xi}^3_0=- \frac{3m}{2r^3} {\xi}^0_3=0 \Rightarrow \fbox{${\xi}^0_3=0$} \]  
\[ {\rho}_{01,23}=0 \,\, \Rightarrow \,\,  R_{01,23}\equiv  {\rho}_{01,23}({\xi}^0_0 + {\xi}^1_1 + {\xi}^2_2 + {\xi}^3_3) + \xi \partial {\rho}_{01,23} = 0  \]    
\[ d_1R_{01,23} + d_2 R_{01,31} +d_3 R_{01,12}= \frac{3m}{2r^3}(d_2{\xi}^0_3 - d_3{\xi}^0_2)=0 \hspace{5mm}mod(\Omega,\Gamma, R)  \]         
a result showing that certain third order CC may be differential consequrences of the Bianchi identities (See [19] for details). Finally, we notice that:   
\[  R_{23,23}\equiv  2{\rho}_{23,23}({\xi}^2_2 + {\xi}^3_3) + \xi \partial {\rho}_{23,23}= 3 m{sin}^2(\theta) {\xi}^1 =0    \]            
and, comparing to the previous computation for $({\xi}^2,{\xi}^3)$, {\it nothing can be said about the generating CC as long as the PP procedure has not been totally achieved with a FI or involutive system}.  \\

 \newpage

\noindent
{\bf 4) KERR METRIC REVISITED}  \\

 Though we shall provide explicitly all the details of the computations involved, we shall change the coordinate system in order to confirm theses results by using computer algebra in a much faster way. The idea is to use the so-called " {\it rational polynomial} " coefficients while setting anew : \\
 \[ (x^0=t, \, x^1=r, \, x^2=c=cos(\theta), \, x^3=\phi)  \Rightarrow dx^2= - sin(\theta)d\theta \Rightarrow (dx^2)^2=(1-c^2)d\theta^2 \]
in order to obtain over the differential field $K= \mathbb{Q}(a,m)(t,r,c,\phi)=\mathbb{Q}(a,m)(x)$: \\
\[  \begin{array}{rcl}
ds^2  &  =  & \frac{{\rho}^2 - mx^1}{{\rho}^2}(dx^0)^2 - \frac{{\rho}^2}{\Delta} (dx^1)^2  -  \frac{{\rho}^2}{1-(x^2)^2} (dx^2)^2  \\
  &    &  - \frac{2a m x^1(1-(x^2)^2) }{{\rho}^2} dx^0dx^3 -  (1-(x^2)^2)((x^1)^2+a^2 + \frac{m a^2 x^1 (1-(x^2)^2)}{{\rho}^2}) (dx^3)^2          
  \end{array}   \]
with now $\Delta= (x^1)^2 - m x^1 +a^2=r^2 - mr + a^2$ and $  {\rho}^2=(x^1)^2 +a^2(x^2)^2=r^2 + a^2c^2$. For a later use, it is also possible to set ${\omega}_{33}= - (1-c^2)((r^2 + a^2)^2 - a^2 ((1-c^2)(a^2 - mr + r^2))/ (r^2 + a^2c^2)$. \\
As this result will be crucially used later on, we have:  \\

\noindent
{\bf LEMMA 4.1}: $det(\omega)= - (r^2 + a^2c^2)^2$ .   \\

\noindent
{\it Proof}: As an elementary result on matrices, we have:   \\
\[    det (\omega)=det\,  \left ( \begin{array}{cccc}
  a & 0 & 0 & e \\
  0 & b & 0 & 0 \\
  0 & 0 & c & 0 \\
  e & 0 & 0 & d
  \end{array}  \right )  
  =   bc \,\,\,det  \, \left ( \begin{array}{cc}
  a & e \\
  e & d
  \end{array} \right )
  = bc(ad-e^2)        \]
with $e={\omega}_{03}= \frac{amx^1(1-(x^2)^2)}{{\rho}^2}$ because $ds^2= ... + 2{\omega}_{03}dx^0dx^3 + ...$ and $det(\omega)$ is thus equal to:   \\
\[  \frac{{\rho}^4}{\Delta (1 - (x^2)^2)} [ - \frac{({\rho}^2-mx^1)}{{\rho}^2}(1-(x^2)^2)((x^1)^2+a^2 + \frac{m a^2 x^1 (1-(x^2)^2)}{{\rho}^2})  - \frac{(amx^1 (1- (x^2)^2))^2}{{\rho}^4}         ]   \]
that is, after division by $(1-(x^2)^2)$ and ${\rho}^4$:  \\
\[   \frac{1}{\Delta } [ - ({\rho}^2-mx^1)({\rho}^2(x^1)^2+{\rho}^2a^2 + m a^2 x^1(1- (x^2)^2) )  - a^2m^2(x^1)^2(1-(x^2)^2)  )   \]       
Finally, after eliminating the last term, we get:  \\
\[  \frac{1}{\Delta} [ -{\rho}^4((x^1)^2 + a^2 ) -{\rho}^2ma^2x^1(1-(x^2)^2)+{\rho}^2mx^1((x^1)^2+a^2)]    \]
that is (Compare to [  ] and [ ]):  \\
\[   \frac{1}{\Delta} [ -{\rho}^4(\Delta + mx^1) +{\rho}^2ma^2x^1(x^2)^2+{\rho}^2m(x^1)^3] =\frac{1}{\Delta}[ - {\rho}^4(\Delta + mx^1)  + 
{\rho}^2mx^1 (a^2(x^2)^2 +(x^1)^2)] = - {\rho}^4  \]
in a coherent way with the result $A(-\frac{1}{A})(- \frac{r^2}{ (1-c^2)}(- r^2 (1-c^2))=-r^4$ obtained for the S metric when $a \rightarrow 0$. For a later use, we have obtained ${\omega}_{00}{\omega}_{33} - ({\omega}_{03})^2= - (1 - c^2)\Delta$. \\
\hspace*{12cm}  Q.E.D.  \\

Contrary to the Schwarzschild metric, the main "trick" for studying the Kerr metric is to take into account that the partition between the zero and nonzero terms will not change if we modify the coordinates, even if, of course, the nonzero terms may change. Meanwhile, we notice that the most important property of the Kerr metric is the off-diagonal term ${\omega}_{t\phi}={\omega}_{\phi t}= -\frac{amsin^2(\theta)}{{\rho}^2}$, that is $\frac{1}{2}$ the coefficient of $dtd\phi$ in the metric $ds^2$ which is indeed $ 2{\omega}_{t\phi}dtd\phi $. We may obtain therefore successively the Killing equations for the Kerr type metric, using sections of jet bundles and writing simply $\xi \partial \omega ={\xi}^r{\partial}_r \omega=
{\xi}^1{\partial}_1\omega + {\xi}^2 {\partial}_2 \omega$ while framing the principal derivative  ${\xi}^j_i$ of ${\Omega}_{ij}$:  \\

\[ R_1\subset J_1(T) \,\,\,  \left\{  \begin{array}{lcl}
{\Omega}_{33} & \equiv & 2( {\omega}_{33}\fbox{${\xi}^3_3$} + {\omega}_{03} {\xi}^0_3 ) + \xi \partial {\omega}_{33}=0 \\
{\Omega}_{23} & \equiv & {\omega}_{33}\fbox{${\xi}^3_2$} + {\omega}_{03}{\xi}^0_2 + {\omega}_{22}{\xi}^2_3 = 0 \\
{\Omega}_{22} & \equiv & 2 {\omega}_{22}\fbox{${\xi}^2_2$} + \xi \partial {\omega}_{22} =0  \\
{\Omega}_{13} & \equiv  & {\omega}_{33}\fbox{${\xi}^3_1$} + {\omega}_{03}{\xi}^0_1 +{\omega}_{11}{\xi}^1_3 =0 \\
{\Omega}_{12} & \equiv  & {\omega}_{22}\fbox{${\xi}^2_1$} +{ \omega}_{11}{\xi}^1_2 =0  \\
{\Omega}_{11} & \equiv  & 2 {\omega}_{11} \fbox{${\xi}^1_1$} + \xi \partial {\omega}_{11} = 0  \\
{\Omega}_{03} & \equiv  & {\omega}_{33}\fbox{${\xi}^3_0$} + {\omega}_{03}({\xi}^0_0 +{\xi}^3_3)  + {\omega}_{00}{\xi}^0_3 + \xi \partial {\omega}_{03} = 0 \\
{\Omega}_{02} & \equiv  & {\omega}_{22}\fbox{${\xi}^2_0$} + {\omega}_{00}{\xi}^0_2  + {\omega}_{03}{\xi}^3_2 = 0 \\
{\Omega}_{01} & \equiv  & {\omega}_{11}\fbox{${\xi}^1_0$} + {\omega}_{00}{\xi}^0_1 + {\omega}_{03}{\xi}^3_1 =0  \\
{\Omega}_{00} & \equiv  & 2 ({\omega}_{00}\fbox{${\xi}^0_0$} + {\omega}_{03} {\xi}^3_0 ) +\xi \partial {\omega}_{00}=0 
\end{array}  \right.  \]
With $mod(\xi)=mod ({\xi}^1,{\xi}^2)$, multiplying ${\Omega}_{33}$ by ${\omega}_{00}$, ${\Omega}_{00}$ by ${\omega}_{33}$ and adding, 
we notice that:  \\
\[   2 {\omega}_{00}{\omega}_{33}({\xi}^0_0 + {\xi}^3_3) +2 {\omega}_{03}({\omega}_{00}{\xi}^0_3 + {\omega}_{33}{\xi}^3_0) + \xi \partial ({\omega}_{00}{\omega}_{33})=0 \]
Similarly, multiplying ${\Omega}_{03}$ by $2 {\omega}_{03}$ ({\it care to the factor} $2$), we get:  \\
\[  2 ({\omega}_{03})^2 ({\xi}^0_0 + {\xi}^3_3) + 2 {\omega}_{03} ( {\omega}_{00}{\xi}^0_3 + {\omega}_{33}{\xi}^3_0)+
\xi\partial ({\omega}_{03})^2=0  \]
Substracting, we obtain therefore the {\it tricky} formula (see the previous Lemma):  \\
\[2({\omega}_{00}{\omega}_{33} - ({\omega}_{03})^2)({\xi}^0_0 + {\xi}^3_3) + \xi \partial ({\omega}_{00}{\omega}_{33} - ({\omega}_{03})^2)=0\]
Substituting, we obtain:  \\
\[ {\omega}_{33}\fbox{${\xi}^3_3$} + {\omega}_{03}{\xi}^0_3=0 \,\, mod(\xi), \,\,\, {\omega}_{33} \fbox{${\xi}^3_0$} + 
{\omega}_{00}{\xi}^0_3 = 0 \,\, mod(\xi), \,\,\, 
  {\omega}_{33}\fbox{${\xi}^0_0$} - {\omega}_{03}{\xi}^0_3=0 \,\, mod(\xi)   \] 
a situation leading to modify ${\Omega}_{33}$, ${\Omega}_{03}$ and ${\Omega}_{00}$, similar to the one found in the Minkowski case with 
${\xi}_{3,3}=0, \,\,\,{\xi}_{0,3} + {\xi}_{3,0}=0, \,\,\, {\xi}_{0,0}=0 \,\, mod(\xi)$ when ${\omega}_{03}=0$. We also obtain with ${\Omega}_{01}$ and ${\Omega}_{13}$:  \\
\[  \fbox{$ \begin{array}{lcl}({\omega}_{00}{\omega}_{33} - ({\omega}_{03})^2)\fbox {${\xi}^0_1$} + {\omega}_{11}({\omega}_{33}{\xi}^1_0 - {\omega}_{03}{\xi}^1_3) &= &0\,\, mod(\xi) \\
  ({\omega}_{00}{\omega}_{33} - ({\omega}_{03})^2) \fbox{${\xi}^3_1$} - 
{\omega}_{11}({\omega}_{03}{\xi}^1_0 - {\omega}_{00}{\xi}^1_3)&= & 0 \,\, mod(\xi)
\end{array}$}  \]
and with ${\Omega}_{02}$ and ${\Omega}_{23}$:  \\
\[  \fbox{$  \begin{array}{lcl} ({\omega}_{00}{\omega}_{33} - ({\omega}_{03})^2)\fbox {${\xi}^0_2$} + 
{\omega}_{22}({\omega}_{33}{\xi}^2_0 - {\omega}_{03}{\xi}^2_3) & = & 0\,\, mod(\xi) \\
  ({\omega}_{00}{\omega}_{33} - ({\omega}_{03})^2) \fbox{${\xi}^3_2$} - 
{\omega}_{22}({\omega}_{03}{\xi}^2_0 - {\omega}_{00}{\xi}^2_3)& = & 0 \,\, mod(\xi) 
\end{array} $} \]

Finally, multiplying ${\Omega}_{22}$ by ${\omega}_{11}$, ${\Omega}_{11}$ by ${\omega}_{22}$ and adding, we finally obtain (see the Lemma again)  \\
 \[   2 ({\omega}_{11}{\omega}_{22})({\xi}^1_1 + {\xi}^2_2) +\xi \partial ({\omega}_{11}{\omega}_{22})=0   \]

Using the rational coefficients belonging to the differential field $K=\mathbb{Q}(m,a)(x^1,x^2)$, the nonzero components of the corresponding Riemann tensor can be found in textbooks. \\
One has the classical orthonormal decomposition:  \\
\[  ds^2=  \frac{\Delta}{{\rho}^2} (dt - a sin^2(\theta)d\phi)^2 - \frac{{\rho}^2}{\Delta} (dr)^2 - {\rho}^2 (d\theta)^2 - \frac{(r^2 + a^2)^2{sin}^2(\theta)}{{\rho}^2}(d\phi - \frac{a}{r^2+a^2} dt)^2   \]
and defining: \\
\[  \left\{  \begin{array}{lcl}
dX^0 & = & \frac{\sqrt{\Delta}}{\rho}(dt - a sin^2(\theta) d\phi)  \\
dX^1 & = & \frac{\rho}{\sqrt{\Delta}}dr=\frac{\rho}{\sqrt{\Delta}}dx^1  \\
dX^2 & = & \rho d\theta = - \frac{\rho}{sin(\theta)}dx^2 \\
dX^3 & = & \frac{(r^2 + a^2)sin(\theta)}{\rho}(d\phi - \frac{a}{r^2 + a^2}dt)
\end{array} \right.  \]
in which the coefficient of $(dt)^2$ is $\frac{\Delta}{{\rho}^2} - \frac{a^2sin^2(\theta)}{{\rho}^2}=1 - \frac{mr}{{\rho}^2}$ while the coefficient of  
$(d\phi)^2$ is $ - (r^2 + a^2 + \frac{mra^2sin^2(\theta)}{{\rho}^2})sin^2(\theta)$ indeed. We have $ds^2=(dX^0)^2 - (dX^1)^2 - (dX^2)^2 - (dX^3)^2$ and make thus the Minkowski metric appearing in a purely alebraic way. We now use the new coordinates $(x^0=t,x^1=r, x^2 =cos(\theta), x^3=\phi)$ and it follows that the conditions ${\xi}^1= 0, {\xi}^2=0$ are invariant under such a change of basis because $dX^1$ and $dX^2$ are respectively proportional to $dx^1$ and $dx^2$. Indeed, as $\omega=\omega (r,\theta)$ and thus $\xi \partial \omega = 0$, the new symbol $g'_1$ of $R'_1=R^{(2)}_1\subset R_1 \subset T^*\otimes T$ while $\rho \in {\wedge}^2T^*Ê\otimes T^*\otimes T$ as mixed tensors. \\
We obtain simpler formulas in the corresponding basis, in particular the $6$ components with only two different indices are proportional to 
$\frac{mr(r^2 - 3 a^2c^2)}{(r^2+a^2c^2)^3} $ while the $3$ components with four different indices are proportional to 
$\frac{amc(3r^2 - a^2c^2)}{(r^2 + a^2c^2)^3}$. \\

In the original rational coordinate system, {\it the main nonzero useful components of the Riemann tensor can only be obtained by means of computer algebra}. For helping the reader with the literature, in particular the book " Computational in Riemann Geometry " wrote by Kenneth R. Koehler that can be found on the net with a free access. We notice that $\omega \rightarrow - \omega$, that is to say changing the sign of the metric, does not change the Christoffel symbols $({\gamma}^k_{ij})$ and the Riemann tensor $( {\rho}^r_{l,ij})$ {\it but changes the sign of} $({\rho}_{kl,ij}={\omega}_{kr}{\rho}^r_{l,ij})$. For this reason, we have adopted the sign convention of this reference for the explicit computation of these later components as the products and quotients used in the sequel will not be changed. \\
We have successively: \\ 
\[ \left\{  \begin{array}{lcl}
{\rho}_{01,01} & = & - \frac{mr(2(r^2 - mr +a^2) +a^2(1-c^2))(r^2- 3 a^2c^2)}{2(r^2 + a^2c^2)^3
(r^2 - mr +a^2)}   \\   \\
{\rho}_{02,02} &  =  &  \frac{mr(r^2 - mr + a^2 + 2 a^2 (1-c^2))(r^2 - 3 a^2c^2)}{2(1-c^2)(r^2 + a^2c^2)^3} \\  \\
{\rho}_{03,03} & = & \frac{ mr (1-c^2)(r^2 - mr +a^2)(r^2 - 3a^2c^2)}{2 (r^2 + a^2c^2)^3}  \\  \\                      
{\rho}_{12,12} & = & - \frac{mr(r^2 - 3a^2c^2)}{ 2 (1-c^2)(r^2 + a^2c^2)(r^2- mr + a^2)}   \\  \\
{\rho}_{13,13} &  =  &  \frac{ - (1-c^2)mr(r^4 - 2a^2c^2 r^2 + 4a^2r^2 - 2 a^4c^2+ 3a^4- 2a^2mr(1-c^2))(r^2 - 3 a^2c^2)}{2(r^2 + a^2c^2)^3(r^2 - mr +a^2)}  \\   \\ 
{\rho}_{23,23} & = & \frac{mr(2r^4 - a^2c^2 r^2 +5a^2r^2 - a^4c^2 + 3 a^4 - a^2mr(1-c^2))(r^2 - 3 a^2c^2)}{2(r^2 + a^2c^2)^3}   \\   \\
{\rho}_{01,23} & = &   \frac{amc(2r^2 - a^2c^2 + 3 a^2)(3r^2 - a^2c^2)}{ 2 (r^2 + a^2c^2)^3}   \\  \\
{\rho}_{02,31} & = & -  \frac{ amc(r^2 - 2 a^2c^2 + 3a^2)(3 r^2 - a^2c^2)}{ 2 (r^2 + a^2c^2)^3}  \\  \\
{\rho}_{03,12} & = & - \frac{amc(3r^2 - a^2c^2)}{2(r^2 + a^2c^2)^2}
\end{array} \right. \]
It must be noticed that we have been able to factorize the six components with only two different indices by $(r^2 - 3a^2c^2)$ and the three components with four different indices by $(3 r^2 - a^2c^2)$, a result not evident at first sight but coherent with the orthogonal decomposition.  \\

After tedious computations, we obtain:   \\
\[  - \frac{{\omega}^{03}}{{\omega}^{11}} {\rho}_{03,03}  =  - (-\frac{{\rho}^2}{\Delta})( \frac{ amr(1-c^2)}{{\rho}^2})(- \frac{1}{(1-c^2)\Delta}) {\rho}_{03,03}  =   - \frac{ a m^2r^2 (1-c^2)(r^2 -3a^2c^2)}{2(r^2 + a^2c^2)^2(r^2 - mr +a^2)}  \]
which is indeed vanishing when $a=0$ for the S metric, both with:  \\
\[   \left\{ \begin{array}{rcl}
   {\rho}_{02,13} + {\rho}_{03,12}  & =  & \frac{3a^3mc(1-c^2)(3r^2 - a^2c^2)}{2(r^2 + a^2c^2)^3}      \\   \\
   {\rho}_{01,23} + {\rho}_{03,21}  & = & \frac{3 amc(r^2 + a^2)(3r^2- a^2c^2)}{2(r^2 + a^2c^2)^3}  
   \end{array}  \right.   \]
   
\[   \left \{  \begin{array}{rcl}
 {\rho}_{02,10}  & =  & \frac{3a^2mc(3r^2 - a^2c^2)}{2(r^2 + a^2c^2)^3}  \\  \\
 {\rho}_{02,32}  & =  & \frac{ a m r (3 r^2 - m r + 3 a^2)(r^2 - 3 a^2c^2) }{ 2 (r^2 + a^2c^2)^3}  \\   \\
 {\rho}_{13,23} & =  & - \frac{ 3a^2 m c(1-c^2)(r^2 + a^2)(3 r^2 - a^2c^2)}{2 (r^2 + a^2c^2)^3}  \\   \\  
 {\rho}_{01,13} &  =  &  \frac{amr(1-c^2)(3r^2 + 3a^2 - 2mr)(r^2 - 3a^2c^2)}{ 2 (r^2 + a^2c^2)^3(r^2 -mr+a^2)}  \\   \\
           &  =  &  \frac{3amr(1-c^2)(r^2-3a^2c^2)}{2(r^2 + a^2c 2^3)} + \frac{am^2r^2(1-c^2)(r^2-3a^2c^2)}{2(r^2+a^2c^2)^3(r^2 - mr +a^2)}
 \end{array} \right.  \]

Introducing the {\it formal Lie derivative} $R=L({\xi}_1)\rho$ and using the fact that $\rho \in {\wedge}^2T^*\otimes T^* \otimes T^*$ is a tensor, 
the system $R^{(2)}_1$ contains the new equations:  \\
\[  R_{kl,ij}\equiv{\rho}_{rl,ij}{\xi}^r_k + {\rho}_{kr,ij}{\xi}^r_l + {\rho}_{kl,rj}{\xi}^r_i + {\rho}_{kl,ir}{\xi}^r_j + {\xi}^r{\partial}_r{\rho}_{kl,ij}=0   \]
Taking into account the original first order Killing equations, we obtain successively:  \\
\[ \left\{ \begin{array}{lcl}
  R_{01,01}& \equiv & 2 {\rho}_{01,01} ({\xi}^0_0 + {\xi}^1_1) + 2 {\rho}_{01,31}{\xi}^3_0 + 2{\rho}_{01,02} {\xi}^2_1 + 
  \xi \partial{\rho}_{01,01} = 0  \\
  R_{02,02} & \equiv & 2 {\rho}_{02,02} ({\xi}^0_0 + {\xi}^2_2) + 2 {\rho}_{02,32}{\xi}^3_0 + 2{\rho}_{01,02}{\xi}^1_2 + 
  \xi \partial {\rho}_{02,02}= 0   \\ 
  R_{03,03} & \equiv & 2 {\rho}_{03,03} ({\xi}^0_0 + {\xi}^3_3) + \xi \partial {\rho}_{03,03} =0    \\
  R_{12,12} & \equiv & 2 {\rho}_{12,12} ({\xi}^1_1 + {\xi}^2_2) + \xi \partial {\rho}_{12,12} =0    \\
  R_{13,13} & \equiv & 2 {\rho}_{13,13} ({\xi}^1_1 + {\xi}^3_3) + 2{\rho}_{13,23}{\xi}^2_1 + 2{\rho}_{13,10}{\xi}^0_3 + 
  \xi \partial {\rho}_{13,13}=0  \\
  R_{23,23} & \equiv & 2 {\rho}_{23,23} ({\xi}^2_2 + {\xi}^3_3) + 2 { \rho}_{13,23}{\xi}^1_2 + 2{\rho}_{20,23}{\xi}^0_3 +      
  \xi \partial {\rho}_{23,23}=0  
    \end{array}  \right.  \]
 and we must add:  \\
 \[  \hspace{-2cm}  \left\{  \begin{array}{lcl}
  R_{01,23} & \equiv & {\rho}_{01,23} ({\xi}^0_0 +{\xi}^1_1 + {\xi}^2_2 + {\xi}^3_3 ) +\xi \partial {\rho}_{01,23}=0   \\
  R_{02,13} & \equiv & {\rho}_{02,13} ({\xi}^0_0 +{\xi}^1_1 + {\xi}^2_2 + {\xi}^3_3 ) + \xi \partial {\rho}_{02,13}=0  \\
  R_{03,12} & \equiv & {\rho}_{03,12} ({\xi}^0_0 +{\xi}^1_1 + {\xi}^2_2 + {\xi}^3_3 ) + \xi \partial {\rho}_{03,12}=0  
  \end{array}  \right. \]
  These linear equations are not linearly independent because: 
  \[   {\rho}_{01,23} + {\rho}_{02,31} + {\rho}_{03,12}=0  \,\,\, \Rightarrow \,\,\,    R_{01,23} + R_{02,31} + R_{03,12}=0  \]
  Also, linearizing while using the Kronecker symbol $\delta$, we get:  \\
  \[     {\omega}_{ir}{\omega}^{kr}={\delta}^k_i \,\, \Rightarrow \,\, {\Omega}^{kl}= - {\omega}^{kr}{\omega}^{ls} {\Omega}_{rs} \]
  Thus, introducing the Ricci tensor  and linearizing, we get:
  \[  \begin{array}{rcl}
   {\rho}_{ij}={\omega}^{rs}{\rho}_{ri,sj}={\omega}^{rs}{\rho}_{ir,js}=0 \,\,  \Rightarrow \,\,R_{ij} & = &  {\omega}^{rs}R_{ri,sj} + {\rho}_{ik,jl} {\Omega}^{kl} \\
            & = &  {\omega}^{rs}R_{ir,js} - {\rho}_{ik,jl} {\omega}^{kr}{\omega}^{ls} {\Omega}_{rs}  \\
            & = &{\rho}_{rj}{\xi}^r_i + {\rho}_{ir}{\xi}^r_j + {\xi}^r {\partial}_r {\rho}_{ij}  \\
            & = & 0
         \end{array}         \]
  It follows that $ - R_{ij}\equiv {\omega}^{rs}R_{ir,sj}=0 \,\,\, mod(\Omega) $ and we have in particular:  \\
  \[  \left\{  \begin{array}{lcl}
  R_{00} &\equiv &{\omega}^{11}R_{01,01} + {\omega}^{22}R_{02,02} + {\omega}^{33}R_{03,03}=0   \,\,\,  mod(\Omega)   \\
  R_{11} &\equiv &{\omega}^{00}R_{01,01} + 2{\omega}^{03}R_{01,31} + {\omega}^{22}R_{12,12}+ {\omega}^{33}R_{13,13}=0 \,\,\, mod(\Omega) \\
  R_{22}& \equiv &{\omega}^{00}R_{02,02} + 2{\omega}^{03}R_{02,32} + {\omega}^{11}R_{12,12}+ {\omega}^{33}R_{23,23}=0 \,\,\, mod(\Omega) \\
  R_{33} &\equiv &{\omega}^{00}R_{03,03} + {\omega}^{11}R_{13,13}+ {\omega}^{22}R_{23,23}=0 \,\,\, mod(\Omega) 
  \end{array} \right.  \]
The first row proves that $R_{03,03}$ is a linear combination of $R_{01,01}$ and $R_{02,02}$. Then, if we want to solve the three other equations 
  with respect to $R_{12,12}$, $R_{13,13}$ and $R_{23,23}$, the corresponding determinant is, up to sign:  \\
\[ det \,\,\,  \left( \begin{array}{lll}
{\omega}^{22} & {\omega}^{33} & 0 \\
{\omega}^{11} & 0 & {\omega}^{33} \\
0  &  {\omega}^{11} & {\omega}^{22}
\end{array} \right) = - 2 \,\, {\omega}^{11}{\omega}^{22}{\omega}^{33}\neq 0   \]
Accordingly, we only need to take into account $R_{01,01}, R_{02,02}$, $R_{01,13}$, $R_{02,23}$.  \\
Similarly, we also obtain:  \\
 \[  \left\{  \begin{array}{lcl}
 R_{01}&\equiv &{\omega}^{22}R_{20,21} + {\omega}^{33}R_{30,31} +{\omega}^{03}R_{30,01}=0 \,\,\, mod(\Omega) \\
 R_{02} & \equiv &{\omega}^{11}R_{01,21} + {\omega}^{33}R_{03,32} + {\omega}^{03}R_{30,02} =0 \,\,\, mod(\Omega)     \\
 R_{03} &\equiv & \fbox {${\omega}^{11}R_{10,13} + {\omega}^{22}R_{20,23}$} + {\omega}^{03}R_{03,03} =0  \,\,\, mod(\Omega)   \\
             &           &    \\
 R_{12} &\equiv & \fbox {${\omega}^{00}R_{01,02} + {\omega}^{33}R_{31,32}$}+ {\omega}^{03}(R_{01,32} + R_{31,02})  =0 \,\,\, 
 mod(\Omega)  \\
 R_{13} & \equiv  & {\omega}^{00}R_{01,03} + {\omega}^{22}R_{21,23}+{\omega}^{03} R_{31,03}=0 \,\,\, mod(\Omega)  \\
 R_{23} &\equiv&{\omega}^{00}R_{02,03} + {\omega}^{11}R_{12,13} + {\omega}^{03}R_{32,03}=0 \,\,\, mod(\Omega)  
 \end{array}  \right.  \]
 where we have to set $R_{01,23}=0, R_{02,13}=0 \,\,  \Rightarrow \,\,  R_{03,12}=0$.  \\
Hence, taking into account $R_{03}=0$, we just need to use $R_{01,01}, R_{02,02}$ and $R_{01,13}$.  \\
However, using the previous lemma, we obtain the formal Lie derivative:  \\
\[   2 \, det(\omega) ( {\xi}^0_0 +{\xi}^1_1 + {\xi}^2_2 + {\xi}^3_3 ) + \xi \partial det(\omega)=0   \]
and thus $\xi \partial ({\rho}_{01,23}/(\sqrt{\mid det(\omega) \mid )} )= 0 $ with $\sqrt{\mid det(\omega) \mid )}= r^2 + a^2 cos^2(\theta)$.  \\
In addition, we have $2 ({\omega}_{11}{\omega}_{22})({\xi}^1_1 + {\xi}^2_2) + \xi \partial  ({\omega}_{11}{\omega}_{22})=0 $ and thus 
$ \xi \partial ({\rho}_{12,12}/ ({\omega}_{11}{\omega}_{22})) = 0 $.  \\
We have also:  \\
\[ 2 ({\rho}_{03,03}{\rho}_{12,12})({\xi}^0_0 + {\xi}^1_1 + {\xi}^2_2 + {\xi}^3_3) + \xi \partial ({\rho}_{03,03}{\rho}_{12,12}) = 0 \,\, 
\Rightarrow \,\,  \xi \partial ( {\rho}_{03,03}{\rho}_{12,12}/det(\omega))=0  \]
The following invariants are obtained successively in a coherent way:   \\
\[  \mid {\rho}_{03,03}{\rho}_{12,12}\mid=  \frac{m^2r^2(r^2 - 3 a^2c^2)^2}{4(r^2 + a^2c^2)^4}\,\, \Rightarrow \,\, 
   \mid {\rho}_{03,03}{\rho}_{12,12}\mid / \mid det(\omega)\mid= (\frac{mr(r^2 - 3 a^2c^2)}{2(r^2 + a^2c^2)^3})^2 \]
\[  {\omega}_{11}{\omega}_{22}= \frac{(r^2 + a^2c^2)^2}{(1-c^2)(r^2 - mr + a^2)} \,\,\, \Rightarrow \,\,\,
\mid {\rho}_{12,12}\mid/({\omega}_{11}{\omega}_{22})= \frac{mr (r^2 - 3 a^2c^2)}{2(r^2+a^2c^2)^3}  \]
However, as $a\in K$, then ${\rho}_{01,23}$ and ${\rho}_{02,13}$ can be both divided by $a$ and we get the new invariant:  \\
\[  {\rho}_{01,23}/{\rho}_{03,12}= \frac{2r^2-a^2c^2+3a^2}{r^2 + a^2c^2} \]
These results are leading to  \fbox{${\xi}^1=0$},  \fbox{${\xi}^2=0$}, thus to \fbox{${\xi}^1_1=0$}, \fbox{${\xi}^2_2=0$} and ${\xi}^0_0 + {\xi}^3_3=0$ after substitution. In the case of the S-metric, only the first invariant can be used in order to find ${\xi}^1=0$. \\

Taking into account the previous result, we obtain the two equations:  \\
\[ \left\{  \begin{array}{l}
 {\rho}_{01,01} ({\xi}^0_0 + {\xi}^1_1) + {\rho}_{01,31}{\xi}^3_0 + {\rho}_{01,02} {\xi}^2_1  = 0  \\
 {\rho}_{02,02} ({\xi}^0_0 + {\xi}^2_2) + {\rho}_{02,32}{\xi}^3_0 + {\rho}_{01,02} {\xi}^1_2   = 0   
\end{array}  \right. \]
Using the fact that we have now:  \\
\[ {\omega}_{22}{\xi}^2_1 + {\omega}_{11}{\xi}^1_2=0 \,\,\, \Leftrightarrow \,\,\, {\omega}^{11}{\xi}^2_1 + {\omega}^{22}{\xi}^1_2=0   \]
we may multiply the first equation by ${\omega}^{11}$, the second by ${\omega}^{22}$ and sum in order to obtain:  \\
\[   ({\omega}^{11}{\rho}_{01,01} + {\omega}^{22}{\rho}_{02,02}){\xi}^0_0 + ({\omega}^{11}{\rho}_{01,31} + 
{\omega}^{22}{\rho}_{02,32}){\xi}^3_0 = 0  \]
Using the previous identity for $R_{03}$, we obtain therefore:  \\
\[    {\omega}^{33} {\rho}_{03,03}{\xi}^0_0 + {\omega}^{03}{\rho}_{03,03}{\xi}^3_0=0 \,\, \Rightarrow \,\, 
{\omega}^{33}{\xi}^0_0 + {\omega}^{03}{\xi}^3_0=0 \,\, \Leftrightarrow \,\, \fbox{ ${\omega}_{03}{\xi}^0_0  - {\omega}_{33}{\xi}^3_0=0$} \]
Taking into account the fact that  ${\xi}^0_0=\frac{{\omega}_{03}}{{\omega}_{33}}{\xi}^0_3, \,\,\, 
{\xi}^3_0= - \frac{{\omega}_{00}}{{\omega}_{33}}{\xi}^0_3$ and substituting, we finally obtain:   \\ 
 \[  ({\omega}_{00}{\omega}_{33} - ({\omega}_{03})^2){\xi}^0_3= 0 \Rightarrow \fbox{${\xi}^0_3=0$},  \fbox{${\xi}^1_2=0$} \,\, \Leftrightarrow \,\,  \fbox{${\xi}^3_0=0$}, \fbox{${\xi}^2_1=0$} , 
 \fbox{${\xi}^0_0=0$}, \fbox{ $ {\xi}^3_3=0 $}  \]
A similar procedure could have been followed by using $R_{13,13}=0, R_{23,23}=0$ and ${\rho}_{33}=0$.  \\

Now, we must distinguish among the $20$ components of the Riemann tensor along with the following tabular where we have to take into account the identity ${\rho}_{01,23} + {\rho}_{02,31} + {\rho}_{03,12}=0$ : \\
\[  \begin{array}{lcccccl}
{\rho}_{01,01} & {\rho}_{01,02} & \fbox{${\rho}_{01,03}$}& \fbox{${\rho}_{01,12}$} & {\rho}_{01,13} & {\rho}_{01,23}\\
{\rho}_{02,02} & \fbox{${\rho}_{02,03}$} & \fbox{${\rho}_{02,12}$}& {\rho}_{02,13}  & & \\
       \\
     \hline \\
{\rho}_{03,03} & {\rho}_{03,12} & \fbox{${\rho}_{03,13}$}& \fbox{${\rho}_{03,23}$}  & {\rho}_{02,23} & \\
{\rho}_{12,12} & \fbox{${\rho}_{12,13}$} & \fbox{${\rho}_{12,23}$} &  & & \\
{\rho}_{13,13} & {\rho}_{13,23} &   &  &  & \\
{\rho}_{23,23} &    &    &  & & 
\end{array}  \]
In this tabular, the vanishing components obtained by computer algebra are put in a box, the nonzero components of the left column do not vanish when $a=0$ and the other components vanish when $a=0$. Also, the 11 ({\it care}) lower components can be known from the 10 upper ones.   \\

Keeping in mind the study of the S-metric and the fact that ${\rho}_{01,03}=0, {\rho}_{03,13}=0, {\rho}_{02,03}=0, {\rho}_{03,13}=0$ while framing the leading terms not vanishing when $a=0$, we get: \\
\[ R_{01,03}\equiv \fbox{${\rho}_{01,01}{\xi}^1_3 + {\rho}_{03,03}{\xi}^3_1 $}+ ( {\rho}_{01,23} + 
{\rho}_{03,21}){\xi}^2_0 + {\rho}_{01,02}{\xi}^2_3 + {\rho}_{01,13}{\xi}^1_0 =0  \]
Then, taking into account the fact that $ {\rho}_{01,12}=0, {\rho}_{02,12}=0, {\rho}_{12,13}=0 $, we obtain similarly:  \\
\[  R_{01,12} \equiv  ({\rho}_{01,32} + {\rho}_{03,12}) {\xi}^3_1 + \fbox{$ {\rho}_{12,21} {\xi}^2_0 + {\rho}_{01,10}{\xi}^0_2 $} +  
+ {\rho}_{01,13}{\xi}^3_2 + {\rho}_{01,02}{\xi}^0_1 =0 \]

The leading determinant does not vanish when $a=0$ because, in this case, all terms are vanishing and we are left with  the two linearly independent framed terms, a result amounting to ${\xi}^1_3=0 \Leftrightarrow {\xi}^3_1=0$ and ${\xi}^0_2=0 \Leftrightarrow {\xi}^2_0=0$ in the case of the S-metric in ([19]). \\
In the case of the K-metric, we may use the relations already framed in order to keep only the four parametric jets $({\xi}^1_3, {\xi}^2_0, {\xi}^1_0, {\xi}^2_3)$ {\it on the right side}. We may also rewrite them as follows:  \\

\[  \fbox{ $ \left\{  \begin{array}{lcl} 
{\omega}^{11}{\xi}^0_1 + {\omega}^{00} {\xi}^1_0 + {\omega}^{03}{\xi}^1_3=0 ,& \,\,\, &
{\omega}^{11}{\xi}^3_1 + {\omega}^{03} {\xi}^1_0 + {\omega}^{33}{\xi}^1_3 =0  \,\,\, \\
{\omega}^{22}{\xi}^0_2 + {\omega}^{00} {\xi}^2_0 + {\omega}^{03}{\xi}^2_3 =0 , &  \,\,\,  &
{\omega}^{22}{\xi}^3_2 + {\omega}^{03} {\xi}^2_0 + {\omega}^{33}{\xi}^2_3 =0
\end{array}  \right. $ } \]
if we use the fact that $ {\omega}^{03} = - {\omega}_{03}/ ({\omega}_{00}{\omega}_{33} - ({\omega}_{03})^2) $ in the inverse metric.  \\

As a byproduct, we are now left with the {\it two} (complicate) equations ${\xi}^1_3 + a ( ... ) =0$ and ${\xi}^2_0 + a ( ... ) =0$ where the dots means linear combinations of $({\xi}^1_0,{\xi}^2_3)$ with coefficients in $K$ and the study of the Killing operator is quite more difficult in the case of the K-metric. Of course, it becomes clear that {\it the use of the formal theory is absolutely necessary as an intrinsic approach could not be achieved if one uses solutions instead of sections}. Indeed the strict inclusion $R'_1=R^{(2)}_1\subset R_1$ cannot be even imagined if one does believe that ${\xi}^1=0, {\xi}^2=0$ brings ${\xi}^1_3=0$ and ${\xi}^2_0=0$. The previous computation could have also be done with $R_{12,23}=0$ and $R_{03,23}=0$ because $R_{02}=0$ and $R_{13}=0$.  \\

The next hard step will be to prove that the other linearized components of the Riemann tensor do not produce {\it any new different} first order 
equation. The main idea will be to revisit the new linearized tabular with:  \\
\[  \begin{array}{lcccccl}
R_{01,01} & R_{01,02} & R_{01,03}& R_{01,12} & R_{01,13} & R_{01,23}\\
R_{02,02} & R_{02,03} & R_{02,12}& R_{02,13}  & & \\
       \\
     \hline \\
R_{03,03} & R_{03,12} & R_{03,13}& R_{03,23}  & R_{02,23} & \\
R_{12,12} & R_{12,13} & R_{12,23} &  & & \\
R_{13,13} & R_{13,23} &   &  &  & \\
R_{23,23} &    &    &  & & 
\end{array}  \]
Puting the leading terms into a box, we have the following formulas:   \\
\[  R_{01,23} + R_{02,31} + \fbox{$R_{03,12}$}= 0  \]
\[  {\omega}^{11}R_{01,13} + \fbox{${\omega}^{22}R_{02,23}$} - ({\omega}^{03}/{\omega}^{33}) ( {\omega}^{11}R_{01,01} + 
{\omega}^{22}R_{02,02})=0  \,\,\, mod(\Omega) \]
\[  {\omega}^{11}R_{01,12} + \fbox{ ${\omega}^{33}R_{03,32}$} + {\omega}^{03}R_{03,02} =0 \,\,\, mod(\Omega)  \]
\[  {\omega}^{00}R_{10,01} + 2{\omega}^{03}R_{10,31} + {\omega}^{22}R_{12,21}+ {\omega}^{33}\fbox{ $R_{13,31}$} =0 \,\,\, 
mod(\Omega)   \]
and so on, allowing to compute the $11$ ({\it care}) lower terms from the $2 + 4 + 4=10$ upper ones.  \\

We have thus the following successive eleven logical inter-relations:  \\
\[   (R_{01,23}, R_{02,13}) \rightarrow R_{03,12}  \]
\[  (R_{01,01}, R_{02,02}, R_{01,13}) \stackrel{(R_{00},R_{11}, R_{22}, R_{33}, R_{03})}{\longrightarrow}
(R_{03,03}, R_{12,12}, R_{13,13}, R_{23,23}, R_{02,23})  \]
\[  (R_{01,02}, R_{01,23}, R_{02,13}) \stackrel{R_{12}}{\longrightarrow} R_{13,23}  \]
\[  (R_{01,03}, R_{02,12}) \stackrel{R_{01}}{\longrightarrow} R_{03,13}  \]
\[  (R_{01,12},R_{02,03}) \stackrel{R_{02}}{\longrightarrow} R_{03,23}  \]
\[  (R_{01,03}, R_{03,13}) \stackrel{R_{13}}{\longrightarrow} R_{12,23}  \]
\[  (R_{02,03}, R_{03,23}) \stackrel{R_{23}}{\longrightarrow}  R_{12,13}  \]

Keeping in mind the four additional equations and their consequences that have been already framed, both with the vanishing components of the Riemann tensor, namely: \\   
\[{\rho}_{01,03}=0, {\rho}_{01,12}=0,{\rho}_{02,03}=0, {\rho}_{02,12}=0, {\rho}_{03,13}=0, {\rho}_{03,23}=0, {\rho}_{12,13}=0, {\rho}_{12,23}=0  \] 
we get successively: \\
\[  R_{01,01}=0, R_{02,02}=0, R_{01,02}=0, R_{01,13}=0, R_{01,23}=0, R_{02,13}=0  \]
As we have already exhibited an isomorphism $ ({\xi}^3_1, {\xi}^0_2,{\xi}^0_1, {\xi}^3_2) \rightarrow ({\xi}^1_3, {\xi}^2_0, {\xi}^1_0, {\xi}^2_3)$, we may use only the later right set of parametric jet components. Using the previous logical relations while framing the leading terms not vanishing {\it a priori} when $a=0$, {\it there is only one possibility to choose four components of the linearized Riemann tensor}, namely:  \\
\[ \left \{ \begin{array}{lcl}
R_{01,03} & \equiv &  \fbox{$({\rho}_{01,01}{\xi}^1_3 + {\rho}_{03,03}{\xi}^3_1) $}+ ( {\rho}_{01,23} + 
{\rho}_{03,21}){\xi}^2_0   + {\rho}_{01,13}{\xi}^1_0  + {\rho}_{01,02}{\xi}^2_3 =0  \\  \\
 R_{03,23} & \equiv &  ({\rho}_{01,23} + {\rho}_{03,21}) {\xi}^1_3 +    \fbox{$ ({\rho}_{03,03} {\xi}^0_2 + {\rho}_{23,23}{\xi}^2_0) $} +  {\rho}_{13,23} { \xi}^1_0 + {\rho}_{02,23} {\xi}^2_3 = 0 \\   \\
 R_{03,13}&  \equiv  & {\rho}_{01,13} {\xi}^1_3  +   {\rho}_{23,13} {\xi}^2_0 +  \fbox{$ ({\rho}_{03,03} {\xi}^0_1 + {\rho}_{13,13}{\xi}^1_0)$} + ({\rho}_{03,12} + {\rho}_{02,13}){ \xi}^2_3 = 0    \\  \\
 R_{02,03} & \equiv & {\rho}_{02,01}{\xi}^1_3 + {\rho}_{02,23}{\xi}^2_0  +  ({\rho}_{03,12} + {\rho}_{02,13}) {\xi}^1_0  +   \fbox{$ ({\rho}_{02,02} {\xi}^2_3 + {\rho}_{03,03} {\xi}^3_2) $}    =0 
 \end{array} \right.  \]  
In order to understand the difficulty of the computations involved, we propose to the reader, as an exercise, to prove " {\it directly} " that the two following relations:  \\ 
\[ R_{02,12} \equiv \fbox{$ ( {\rho}_{12,12}{\xi}^1_0 + {\rho}_{02,02} {\xi}^0_1) $} + ({\rho}_{02,13} + {\rho}_{03,12}){\xi}^3_2 + {\rho}_{02,10}{\xi}^0_2 + {\rho}_{02,32} {\xi}^3_1 = 0  \]
\[  R_{12,13} \equiv \fbox{$({\rho}_{13,13} {\xi}^3_2 + {\rho}_{12,12}{\xi}^2_3)$} + ({\rho}_{03,12} + {\rho}_{02,13}){ \xi}^0_1 + {\rho}_{32,13} {\xi}^3_1 + {\rho}_{10,13} {\xi}^0_2 = 0  \]
are only linear combinations of the previous ones $mod(\Omega)$.  \\

We are facing two technical problems " spoilting ", {\it in our opinion}, the use of the K metric:   \\

\noindent
$\bullet$ \,\, With ${\omega}^{-1}$ in place of $\omega$, we have ${\omega}^{11}{\xi}^3_1= - {\omega}^{33} {\xi}^1_3 + ...$ and the leading term of $R_{01,03}$ becomes proportional to $({\omega}^{11}{\rho}_{01,01} - {\omega}^{33}{\rho}_{03,03}){\xi}^1_3 + ...$ with a {\it wrong sign} indeed that cannot allow to use $R_{00}$. A similar comment is valid for the four successive leading terms.  \\

\noindent
$\bullet$ \,\, In addition, we also discover the summation ${\rho}_{01,23} + {\rho}_{03,21}$ in $R_{01,03}$ with a {\it wrong sign} indeed that cannot allow to introduce ${\rho}_{02,31}$ as one could hope. A similar comment is valid for the four successive summations.  \\

Nevertheless, we obtain the following unexpected formal linearized result that will be used in a crucial intrinsic way for finding out the generating second order and third order CC:   \\

\noindent
{\bf THEOREM  4.2}: The rank of the previous system with respect to the four jet coodinates $({\xi}^1_3, {\xi}^2_0, {\xi}^1_0,{\xi}^2_3)$ is equal to $2$, for both the S and K metrics. We obtain in particular the two striking identities:  \\
\[  \fbox{ $  R_{03,13} + a(1-c^2) R_{01,03}=0, \,\,\,\,\,\,   R_{02,03} + \frac{a}{(r^2+a^2)} R_{03,23} $ } \]   

\noindent
{\it Proof}: In the case of he S metric with $a=0$, only the framed terms may not vanish and, denoting by " $\sim$ " a linear proportionality, we have already obtained $mod(j_2(\Omega))$:  \\
\[     R_{01,03} \sim {\xi}^1_3,\,\,  R_{03,23} \sim {\xi}^0_2, \,\, R_{02,03}=0, \,\, R_{03,13}=0  \]
Hence, the rank of the system with respect to the $4$ parametric jets $({\xi}^1_3, {\xi}^2_0, {\xi}^1_0, {\xi}^2_3)$ just drops to $2$ and this fact confirms the existence of the $5$ additional first order equations obtained, as we saw, after two prolongations.  \\

In the case of the K metric with $a\neq 0$, the study is much more delicate. \\
With $a^0=1$, the coefficients of the $4 \times  4$ metric of the previous system on the basis of the above parametric jets are proportional to the symmetric matrix:\\
\[ \left (  \begin{array}{cccc}
1 & a & a & a^2  \\
a & 1 & a^2 & a  \\
a & a^2 &a^2 & a^3  \\
a^2 & a & a^3 &  a^2  
\end{array} \right )   \]
Indeed, we have successively for the common factor $ - a (1- c^2)$:  \\
\[  \left\{ \begin{array}{lccccl}
Row \,\,  1 & {\xi}^1_3 & \rightarrow & {\rho}_{01,01} - \frac{{\omega}^{33}}{{\omega}_{11}}{\rho}_{03,03} & =  &  
 - \frac{3 m r(r^2 - 3a^2c^2)}{2(r^2 + a^2c^2)^3}  \\
Row \,\,3 &  {\xi}^1_3 & \rightarrow &  {\rho}_{01,13}  &  =  & \frac{3amr(1-c^2)(r^2-3a^2c^2)}{2 (r^2 + a^2c^2)^3}                                                                                                                                                                                                                 
\end{array}\right.  \]

\[  \left\{ \begin{array}{lccccl}
Row \,\,  1 & {\xi}^2_0 & \rightarrow & {\rho}_{01,23} + {\rho}_{03,21} & =  &  
  \frac{3 a m c (r^2 + a^2)(3 r^2 - a^2c^2)}{2(r^2 + a^2c^2)^3}  \\
Row \,\,3 &  {\xi}^2_0 & \rightarrow &  {\rho}_{23,13} &  =  & - \frac{3a^2mc(1-c^2)(r^2 + a^2)(3r^2-a^2c^2)}{2 (r^2 + a^2c^2)^3}                                                                                                                                                                                                                 
\end{array}\right.  \]

\[  \left\{ \begin{array}{lccccl}
Row \,\,  1 & {\xi}^1_0 & \rightarrow & {\rho}_{01,13} - \frac{{\omega}^{03}}{{\omega}_{11}}{\rho}_{03,03} & =  &  
  \frac{3 a m r (1 - c^2)(r^2 - 3a^2c^2)}{2(r^2 + a^2c^2)^3}  \\
Row \,\,3 &  {\xi}^1_0 & \rightarrow &  {\rho}_{13,13} - \frac{{\omega}^{03}}{{\omega}_{11}} {\rho}_{03,03} &  =  & 
- \frac{3a^2mr(1-c^2)^2(r^2-3a^2c^2)}{2 (r^2 + a^2c^2)^3}                                                                                                                                                                                                                 
\end{array}\right.  \]

\[  \left\{ \begin{array}{lccccl}
Row \,\,  1 & {\xi}^2_3 & \rightarrow & {\rho}_{01,02}  & =  &  - \frac{3 a^2 m c(3r^2 - a^2c^2)}{2(r^2 + a^2c^2)^3}  \\
Row \,\,3 &  {\xi}^2_3 & \rightarrow &  {\rho}_{03,12} + {\rho}_{02,13} &  =  & 
\frac{3a^3mc(1-c^2)(3r^2-a^2c^2)}{2 (r^2 + a^2c^2)^3}                                                                                                                                                                                                                 
\end{array}\right.  \]

and similarly for the common factor $ - \frac{a}{(r^2 + a^2)}$:  \\
\newpage
\[  \left\{ \begin{array}{lccccl}
Row \,\,  2 & {\xi}^1_3 & \rightarrow & {\rho}_{01,23} + {\rho}_{03,021} & =  &  
 - \frac{3 amc(r^2 + a^2)(3r^2 - a^2c^2)}{2(r^2 + a^2c^2)^3}  \\
Row \,\,4 &  {\xi}^1_3 & \rightarrow &  {\rho}_{02,01}  &  =  & - \frac{3a^2mc(3r^2-a^2c^2)}{2 (r^2 + a^2c^2)^3}                                                                                                                                                                                                                 
\end{array}\right.  \]

\[  \left\{ \begin{array}{lccccl}
Row \,\,  2 & {\xi}^2_0 & \rightarrow & {\rho}_{23,23} - \frac{{\omega}^{00}}{{\omega}^{22}} {\rho}_{03,03} & =  &  
  \frac{3  m r (r^2 + a^2)^2( r^2 - 3 a^2c^2)}{2(r^2 + a^2c^2)^3}  \\
Row \,\,4 &  {\xi}^2_0 & \rightarrow &  {\rho}_{02,23} - \frac{{\omega}^{03}}{{\omega}^{22}}{\rho}_{03,03}&  =  & 
- \frac{3amr(r^2 + a^2)(r^2 - 3a^2c^2)}{2 (r^2 + a^2c^2)^3}                                                                                                                                                                                                                 
\end{array}\right.  \]

\[  \left\{ \begin{array}{lccccl}
Row \,\,  2 & {\xi}^1_0 & \rightarrow & {\rho}_{13,23} & =  & - \frac{3 a ^2m c (1 - c^2)(r^2 + a^2)(3r^2 - a^2c^2)}{2(r^2 + a^2c^2)^3}  \\
Row \,\,4 &  {\xi}^1_0 & \rightarrow &  {\rho}_{03,12} + {\rho}_{02,13} &  =  & 
 \frac{3a^3mc(1-c^2)(3r^2-a^2c^2)}{2 (r^2 + a^2c^2)^3}                                                                                                                                                                                                                 
\end{array}\right.  \]

\[  \left\{ \begin{array}{lccccl}
Row \,\,  2 & {\xi}^2_3 & \rightarrow & {\rho}_{02,23} - \frac{{\omega}^{03}}{{\omega}^{22}}{\rho}_{03,03} & =  & 
 - \frac{3 a m r(r^2+a^2)(r^2 - 3a^2c^2)}{2(r^2 + a^2c^2)^3}  \\
Row \,\,4 &  {\xi}^2_3 & \rightarrow &  {\rho}_{02,02} - \frac{{\omega}^{33}}{{\omega}^{22}} {\rho}_{03,03} &  =  & 
\frac{3a^2mr(r^2-3a^2c^2)}{2 (r^2 + a^2c^2)^3}                                                                                                                                                                                                                 
\end{array}\right.  \]

We do not believe that such a purely computational mathematical result, {\it though striking it may look like}, could have any useful physical application and this comment will be strengthened by the next theorem provided at the end of this section.  \\
\hspace*{12cm}   Q.E.D.   \\

\noindent
{\bf COROLLARY 4.3}: The Killing operator for the K metric has $14$ generating second order CC.  \\ 

\noindent
{\it Proof}: According to the previous theorem, we have $dim(R^{(2)}_1)= dim(R_3)=4$ as we can choose the $4$ parametric jets $({\xi}^0, {\xi}^3, {\xi}^1_0, {\xi}^2_3)$ and $g_3=0$. Using the {\it introductory diagram} with $n=4, q=1, r=2, E=T$ and thus $dim(J_2(F_0))- dim(J_3(T))= 150 - 140 = 10$, we obtain at once $dim(Q_2)=10 + dim(R^{(2)}_1)=14$ in a purely intrinsic way. We may thus start afresh with the new first order system $R'_1= R^{(2)}_1 \subset R_1 \subset J_1(T)$ obtained from $R_1$ after $2$ prolongations.  \\
\hspace*{12cm}  Q.E.D.  \\

Finally, we know from ([7-9],[16],[18],[19]) that if $R_q\subset J_q(T)$ is a system of infinitesimal Lie equations, then we have the algebroid bracket and its link with the prolongation/projection (PP) procedure:  \\
\[  [R_q,R_q]\subset R_q \Rightarrow [R^{(s)}_{q+r}, R^{(s)}_{q+r}] \subset R^{(s)}_{q+r}, \forall q,r,s \geq 0  \]
It follows that $R'_1=R^{(2)}_1={\pi}^3_1(R_3)$ is such that $[R'_1,R'_1]\subset R'_1$ with $dim(R'_1)= 20-16=4$ because we have obtained a total of $6$ {\it new different} first order equations. Using the first general diagram of the Introduction, we discover that the operator defining $R_1$ has $10+4=14$ CC of order $2$, a result obtained {\it totally independently of any specific GR technical object} like the {\it Teukolski scalars} or the {\it Killing-Yano tensors} introduced in ([1-6]).  \\

It remains to make one more prolongation in order to study $R^"_1= R^{(3)}_1={\pi}^4_1 (R_4) \subset R'_1 \subset R_1$ with strict inclusions in order to sudy the third order CC for $\Omega$ already described for the Schwarzschild metric in ([19]). \\
We have {\it on sections} ({\it care}) the $16$ (linear) equations of $R'_1$ as follows:\\
\[ R'_1=R^{(2)}_1 \subset J_1(T) \, \left\{   \begin{array}{lcl} 
{\xi}^1=0,{\xi}^2=0  &  \Rightarrow & \fbox{ $ {\omega}_{00}{\xi}^0_1 + {\omega}_{03}{\xi}^3_1$} + {\omega}_{11}{\xi}^1_0 =0 ,  \,\, {\xi}^1_1=0, \,\,{\xi}^2_2=0  \\ 
{\xi}^1_2=0  &  \Rightarrow & {\xi}^2_1=0  \\
{\xi}^1_3 + ... =0  & \Rightarrow & \fbox{ $ {\omega}_{03}{\xi}^0_1 + {\omega}_{33}{\xi}^3_1$} + {\omega}_{11}{\xi}^1_3 =0   \\
{\xi}^2_0 + ... =0  & \Rightarrow & \fbox{ $ {\omega}_{00}{\xi}^0_2 +  {\omega}_{03}{\xi}^3_2 $}+ 
{\omega}_{22}{\xi}^2_0 =0 ,  \\
    &  &  \fbox{ $ {\omega}_{03}{\xi}^0_2 +{\omega}_{33}{\xi}^3_2 $}+ {\omega}_{22}{\xi}^2_3=0 \\
{\xi}^0_3=0  & \Rightarrow & {\xi}^3_0=0, \,\, {\xi}^0_0=0, \,\,{\xi}^3_3=0  
\end{array} \right.   \]
and we may choose only the $2$ parametric jets $({\xi}^1_0,{\xi}^2_3)$ among $ ({\xi}^1_0, {\xi}^1_3,{\xi}^2_0, {\xi}^2_3)$ to which we must add $({\xi}^0,{\xi}^3)$ {\it in any case} as they are not appearing in the Killing equations and their prolongations. \\
The system is {\it not} involutive because it is finite type with $g'_2=0$ and $g'_1$ cannot be thus involutive.   \\

Taking therefore into account that the metric only depends on $(x^1=r,x^2=cos(\theta))$ we obtain {\it after three prolongations} the first order system:  \\
\[R^"_1 \subset R'_1\subset  R_1 \subset J_1(T) \,\,\,  \left\{  \begin{array}{lcl}
 {\xi}^3_3   & = & 0 \\
 {\xi}^2_3 & = & 0  \\
 {\xi}^1_3 & = & 0  \\
 {\xi}^0_3 & = & 0  \\
{\xi}^3_2  & = & 0 \\
 {\xi}^2_2  & = & 0  \\
 {\xi}^1_2 & = & 0  \\
 {\xi}^0_2 & = & 0 \\
 {\xi}^3_1  & = & 0 \\
 {\xi}^2_1 & = & 0  \\
 {\xi}^1_1 & = & 0  \\
 {\xi}^0_1  & = & 0  \\
 {\xi}^3_0    & = & 0 \\
 {\xi}^2_0  & = & 0 \\
 {\xi}^1_0  & = & 0  \\
 {\xi}^0_0   & = & 0 \\
 {\xi}^2 & = & 0  \\
 {\xi}^1 & = & 0 
  \end{array} \right. \fbox{ $\begin{array}{llll}
  0 & 1 & 2 & 3  \\
  
    0 & 1 & 2 & 3  \\
     0 & 1 & 2 & 3  \\
    0 & 1 & 2 & 3  \\
  0 & 1 & 2 & \bullet  \\
  0 & 1 & 2 & \bullet  \\
    0 & 1 & 2 & \bullet  \\
  0 & 1 & 2 & \bullet  \\
  0 & 1  &  \bullet &  \bullet \\
  0 & 1  &  \bullet &  \bullet \\
  0 & 1  &  \bullet &  \bullet \\
  0 & 1  &  \bullet &  \bullet \\
0 & \bullet & \bullet & \bullet \\
0 & \bullet & \bullet & \bullet \\
0 & \bullet & \bullet & \bullet \\
0 & \bullet & \bullet & \bullet \\
\bullet & \bullet & \bullet & \bullet \\
\bullet & \bullet & \bullet & \bullet 
  \end{array} $ }   \]   \\ 
{\it Surprisingly and contrary to the situation found for the S metric}, we have now a trivially involutive first order system with only solutions $({\xi}^0=cst, {\xi}^1=0, {\xi}^2=0, {\xi}^3=cst)$. However, the difficulty is to know what second members must be used along the procedure met for all the motivating examples. In particular, we have again identities to zero like $d_0{\xi}^1 - {\xi}^1_0=0, d_3{\xi}^2 - {\xi}^2_3=0$ or, {\it equivalently}, $ d_3{\xi}^1 - {\xi}^1_3=0, d_0{\xi}^2 - {\xi}^2_0=0 $ and thus $4$ third order CC coming from the $4$ following components of the Spencer operator:  \\
\[  d_1{\xi}^1 - {\xi}^1_1=0,\,\, d_2{\xi}^1 - {\xi}^1_2=0, \,\,  d_1{\xi}^2 - {\xi}^2_1=0, \,\, d_2{\xi}^2 - {\xi}^2_2=0  \] 
a result that cannot be even imagined from ([1-6]). Of course, proceeding like in the motivating examples, we must substitute in the right members the values obtained from $j_2(\Omega)$ and set for example ${\xi}^1_1= - \frac{1}{2{\omega}_{11}}\xi \partial {\omega}_{11}$ while replacing ${\xi}^1$ and ${\xi}^2$ by the corresponding linear combinations of the Riemann tensor already obtained for the right members of the two zero order equations. \\

Using one more prolongation, all the {\it sections} ({\it care again}) vanish but ${\xi}^0$ and ${\xi}^3$, a result leading to $dim(R^"_1)=2$ in a coherent way with the only nonzero Killing vectors $\{ {\partial}_t, {\partial}_{\phi} \}$. We have indeed:  \\
\[   \fbox{$  {\xi}^1_0 =0$}  \Rightarrow   {\xi}^3_1=0 \Rightarrow {\xi}^0_1=0 , \hspace{1cm}
\fbox{ $  {\xi}^2_0=0$}   \Rightarrow  {\xi}^3_2=0  \Rightarrow {\xi}^2_3=0  \]

 Like in the case of the S metric, $R_3$ is {\it not} involutive but $R_4$ is involutive. However, contrary to the S metric with $g^"_1\neq 0$, now $g^"_1=0$ for the K metric and $R^"_1$ is trivially involutive with a full Janet tabular having $16$ rows of first order jets and $2$ rows of zero order jets.   \\

\noindent
{\bf REMARK 4.4}: We have in general ([7],[10] p 339, 345):  \\
\[  R^{(s)}_{q+r}= {\pi}^{q+r+s}_{q+r}(R_{q+r+s})={\pi}^{q+r+s}_{q+r}(J_r(R_{q+s})\cap J_{q+r+s}(E))\]
\[\subseteq J_r({\pi}^{q+s}_q)((R_{q+s})\cap J_{q+r}(E)=J_r(R^{(s)}_q)\cap J_{q+r}(E) = {\rho}_r( R^{(s)}_q)  \]
that is, in our case $R^{(2)}_2 \subseteq {\rho}_1(R^{(2)}_1)$. However, we have indeed the equality $R^{(2)}_2={\rho}_1(R^{(2)}_1)$ even if the conditions of Theorem 1.1 are not satisfied because $g'_1$ is not $2$-acyclic. Indeed, the Spencer map $\delta: {\wedge}^2T^*\otimes g'_1 \rightarrow {\wedge}^3T^*\otimes T$ is not injective and we let the reader check as an exercise that its kernel is generated by $\{  v^0_{1,01}, v^3_{2,23}\}$ and the Spencer $\delta$-cohomology is such that $dim(H^2_1(g'_1)=2\neq 0$ because the cocycles are defined by the equations $   v^k_{i,jr} + v^k_{j,ri} + v^k_{r,ij}=0$. Hence, {\it contrary to what could be imagined}, the major difference between the S and K metrics is not at all the existence of off-diagonal terms but rather the fact that $R^"_1$ {\it is not involutive} with $g^"_1\neq 0$ for the S metric while $R^"_1$ {\it is involutive} with $g^"_1=0$ for the K metric. This is the reason for which {\it one among the four third order CC must be added with two prolongations} for the S metric while {\it the four third order CC are obtained in the same way from the Spencer operator} for the K metric. Of course no classical approach can explain this fact which is lacking in ([1-4]).   \\

The following result even questions the usefulness of the whole previous approach:  \\ 

\noindent
{\bf THEOREM 4.5}: The operator $Cauchy = ad(Killing)$ admits a minimum parametrization by the operator $Airy = ad(Riemann)$ with $1$ potential when $n=2$, found in $1863$. It admits a canonical self-adjoint parametrization by the operator $Beltrami = ad(Riemann)$ with $6$ potentials when $n=3$, found in $1892$ and modified to a mimimum parametrization by the operator $Maxwell$ with $3$ potentials, found in $1870$. More generally, it admits a canonical parametrization by the operator $ad(Riemann)$ with $n^2(n^2 - 1)/12$ potentials that can be modified to a relative parametrization by $ad(Ricci)$ with $n(n + 1)/2$ potentials which is nevertheless not minimum when $n\geq 4$, found in $2007$. In all these cases, {\it the corresponding potentials have nothing to do with the perturbation of the metric}. Such a result is also valid for any Lie group of transformations, in particular for the conformal group in arbitrary dimension.  \\

\noindent
{\it Proof}: We provide successively the explicit corresponding parametrizations:  \\

\noindent
$\bullet$ \fbox{$n=2$}: Multiplying the linearized Riemann operator by a test function $\phi$ and integrating by parts, we obtain ({\it care to the factor 2 involved}):  \\

\[  \phi (d_{22}{\Omega}_{11} - \fbox{2} d_{12}{\Omega}_{12} + d_{11}{\Omega}_{22})=(d_{22}\phi{\Omega}_{11} - \fbox{2} d_{12}\phi{\Omega}_{12}+d_{11}\phi{\Omega}_{22}) + div(...) \]

\[ {\sigma}^{ij}={\sigma}^{ji}\,\,\,  \Rightarrow \,\,\,  {\sigma}^{ij}{\Omega}_{ij}= {\sigma}^{11}{\Omega}_{11} + \fbox{2} {\sigma}^{12}{\Omega}_{12}+ {\sigma}^{22}{\Omega}_{22}  \]

\noindent
$Cauchy$ operator\,\,\, \fbox{$d_1{\sigma}^{11} + d_2{\sigma}^{12}=f^1, \,\,\, d_1{\sigma}^{21} + d_2{\sigma}^{22}=f^2$}  \\

\noindent
$Airy$ operator \,\,\, \fbox{ ${\sigma}^{11}=d_{22}\phi, \,\,\,{\sigma}^{12}={\sigma}^{21}= - d_{12} \phi, \,\,\, {\sigma}^{22}=d_{11}\phi $}\\

\[  \begin{array}{rcccccccl}
 &  &  \xi & \longrightarrow & \Omega & \longrightarrow  & R  & \longrightarrow &  0  \\
& & 2 & \stackrel{Killing}{\longrightarrow}& 3 & \stackrel{Riemann}{\longrightarrow} & 1 & \longrightarrow & 0  \\
 & & &  &  &  &  &  &  \\
0& \longleftarrow & 2 & \stackrel{Cauchy}{\longleftarrow} & 3 & \stackrel{Airy}{\longleftarrow} & 1 &  &  \\
0  &  \longleftarrow & f  & \longleftarrow &  \sigma  & \longleftarrow &  \phi  &  &  
\end{array}   \]   \\

\noindent
$\bullet$ \fbox{n=3}  We now present the original $Beltrami$ parametrization:  \\

\noindent
  \[ \large { \left\{  \begin{array}{c}
{\sigma}^{11}  \\  {\sigma}^{12} \\ {\sigma}^{13} \\ {\sigma}^{22} \\ {\sigma}^{23} \\  {\sigma}^{33}
\end{array}  \right\} =   \left\{  \begin{array}{cccccc}
0 & 0 & 0 & d_{33} & -2d_{23} & d_{22}  \\
0 & -d_{33} & d_{23} & 0 & d_{13} & - d_{12}  \\
0 & d_{23} & - d_{22} & - d_{13} & d_{12} & 0  \\
d_{33} & 0 & - 2 d_{13} & 0  &  0  & d_{11}   \\
- d_{23} & d_{13} & d_{12} & 0 & - d_{11} & 0  \\
d_{22} &- 2d_{12}& 0 & d_{11} &  0 & 0 
\end{array}  \right\}  \left\{\begin{array}{c} {\phi}_{11}\\{\phi}_{12}\\ {\phi}_{13}\\ {\phi}_{22} \\ {\phi}_{23} \\ {\phi}_{33}
\end{array}  \right\} } \]  \\ 

\noindent
which  does not seem to be self-adjoint but is such that $d_r {\sigma}^{ir}=0 $. Accordingly, the $Beltrami$ parametrization of the {\it Cauchy} operator for the stress is nothing else than the formal adjoint of the {\it Riemann} operator. However, modifying slightly the rows, we get the new operator matrix:  \\

\[ \large { \left\{  \begin{array}{c}
{\sigma}^{11}  \\  2{\sigma}^{12} \\ 2 {\sigma}^{13} \\ {\sigma}^{22} \\ 2{\sigma}^{23} \\  {\sigma}^{33}
\end{array}  \right\} =   \left\{  \begin{array}{cccccc}
0 & 0 & 0 & d_{33} & -2d_{23} & d_{22}  \\
0 & - 2 d_{33} & 2d_{23} & 0 & 2d_{13} & - 2d_{12}  \\
0 & 2d_{23} & - 2d_{22} & - 2d_{13} & 2d_{12} & 0  \\
d_{33} & 0 & - 2 d_{13} & 0  &  0  & d_{11}   \\
- 2d_{23} & 2d_{13} & 2d_{12} & 0 & - 2 d_{11} & 0  \\
d_{22} &- 2d_{12}& 0 & d_{11}&  0 & 0 
\end{array}  \right\}  \left\{\begin{array}{c} {\phi}_{11}\\{\phi}_{12}\\ {\phi}_{13}\\ {\phi}_{22} \\ {\phi}_{23} \\ {\phi}_{33}
\end{array}  \right\} } \]     

\noindent
which is indeed {\it self-adjoint}. Keeping $({\phi}_{11}=A, {\phi}_{22}=B, {\Phi}_{33}=C)$ with $({\phi}_{12}=0,{\phi}_{13}=0, {\phi}_{23}=0)$, we obtain the $Maxwell$ parametrization:  \\
\[ \large { \left\{  \begin{array}{c}
{\sigma}^{11}  \\  {\sigma}^{12} \\  {\sigma}^{13} \\ {\sigma}^{22} \\ {\sigma}^{23} \\  {\sigma}^{33}
\end{array}  \right\} =   \left\{  \begin{array}{ccc}
0 & d_{33} & d_{22}  \\
0 &  0 &  -d_{12}  \\
0 & - d_{13} & 0  \\
d_{33} &  0  & d_{11}   \\
- d_{23} &  0 & 0  \\
d_{22} & d_{11}&  0 
\end{array}  \right\}  \left\{\begin{array}{c} A\\ B \\ C
\end{array}  \right\} } \]  
which is minimum because $n(n-1)/2=3$. However, the corresponding operator is FI because it is homogeneous but {\it it is not evident at all} to prove that it is also involutive as we must look for $\delta$-regular coordinates (See [15] for the technical details). \\

\noindent
$\bullet$ \fbox{ $ n\geq 4 $} This is far more complicate and {\it we do believe that it is not possible to avoid using differential homological algebra}, in particular {\it extension modules}. As we found it already in many books ([8],[9],[16],[18]) or papers ([12],[13],[21-23]), the linear Spencer sequence is (locally) isomorphic to the tensor product of a Poincar\'{e} type sequence for the exterior derivative by a Lie algebra ${\cal{G}}$ with $dim({\cal{G}})\leq n(n+1)/2$ equal to the dimension of the largest group of invariance of the metric involved. When $n=4$, this dimension is $10$ for the M-metric, $4$ for the S-metric and $2$ for the K-metric. As a byproduct, the adjoint sequence roughly just exchanges the exterior derivatives up to sign and one has for example, when $n=3$, the relations $ad(grad)= - div, ad((div)= - grad$. It follows that, if $D_2$ generates the CC of $D_1$, then $ad(D_2)$ is parametrizing $ad(D_1)$, {\it a fact not evident at all}, even  when $n=2$ for the {\it Cosserat couple-stress equations} {\it exactly} described by $ad(D_1)$ ([12]). Passing to the differential modules point of view with the ring (even an integral domain) $D=K[d_1,...,d_n]=K[d]$ of differential operators with coefficients in a differential field $K$, this result amounts to say that $ext^1_D(M,D)=ext^1(M)=0$. As it is known that such a result does not depend on the differential resolution used or, {\it equivalently}, on the differential sequence used, if ${\cal{D}}_1$ generates the CC of ${\cal{D}}$ in the Janet sequence, then $ad({\cal{D}}_1)$ is parametrizing $ad({\cal{D}})$ and this result is still true even if ${\cal{D}}$ is not involutive. In such a situation, which is the one considered in this paper, the $Killing$ operators for the M-metric, the S-metric and the K-metric are such that, {\it whatever are the generating CC} ${\cal{D}}_1$ (second order for the M-metric, a mixture of second and third order for the S-metric and K-metric), then $ad({\cal{D}}_1)$ is, {\it in any case}, parametrizing the Cauchy operator $ad({\cal{D}})$ for {\it any} ${\cal{D}}: T \rightarrow S_2T^*: \xi \rightarrow {\cal{L}}(\xi)\omega$. Once more, {\it the central object is the group, not the metric}. The same results are also valid for any  Lie group of transformations, in particular for the conformal group in arbitrary dimension, even if the operator ${\cal{D}}_1$ is of order $3$ when $n=3$ as we shall see below ([16],[20-23]).  \\
\hspace*{12cm}   Q.E.D.   \\

\noindent
{\bf REMARK 4.6}: Accordingly, the situation met today in GR cannot evolve as long as people will not acknowledge the fact that the components of the Weyl tensor are {\it similarly} playing the part of torsion elements (the so-called {\it Lichnerowicz waves} in [17]) for the equations $Ricci=0$, a result only depending on the group structure of the conformal group of space-time that brings the canonical splitting $  Riemann = Weyl \oplus Ricci $ without any reference to a backgroung metric as it is usually done ([8],[9],[15],[18],[21-23]). It is an open problem to know why one may sometimes find a SELF-ADJOINT OPERATOR. It is such a confusion that led to introduce the so-called $Einstein$ parametrizing operator ([17]). \\
  
\noindent
{\bf EXAMPLE 4.7}: ({\it  Weyl tensor for n=3 and euclidean metric}) We proved in ([16], p 156-158) and more recently in ([15],[22],[23]) that, for $n=3$, the natural "geometric object " corresponding to the Weyl tensor is no longer described by a second order differential operator but by a third order differential operator $\hat{\cal{D}}_1$ with first order CC $\hat{\cal{D}}_2$ in the differential sequence: \\
\[          0 \longrightarrow \hat{\Theta} \longrightarrow 3  \underset 1 {\stackrel{\hat{\cal{D}}}{\longrightarrow}} 5 \underset 3{\stackrel{\hat{\cal{D}}_1}{\longrightarrow}} 5 \underset 1{\stackrel{\hat{\cal{D}}_2}{\longrightarrow}} 3 \longrightarrow 0  \]
corresponding to the differential sequence of $D$-modules:  \\
\[          0  \longrightarrow D^3  \underset 1 {\stackrel{\hat{\cal{D}}_2}{\longrightarrow}} D^5 \underset 3{\stackrel{\hat{\cal{D}}_1}{\longrightarrow}} D^5 \underset 1{\stackrel{\hat{\cal{D}}}{\longrightarrow}} D^3 \stackrel{p}{\longrightarrow}  \hat{M} \longrightarrow 0  \]
where $p$ is the canonical residual projection. The true reason is that the symbol ${\hat{g}}_1$ of $\hat{\cal{D}}$ is finite type with second prolongation ${\hat{g}}_3=0$ while its first prolongation ${\hat{g}}_2$ is {\it not} $2$-acyclic. It is important to notice that the operators are acting on the left on column vectors in the upper sequence but on the right on row vectors in the lower sequence though we have {\it in any case} the identities $\hat{\cal{D}}_1 \circ \hat{\cal{D}}=0$ and $\hat{\cal{D}}_2\circ \hat{\cal{D}}_1=0$.\\
Of course, these operators can be obtained by using computer algebra like in ([16], Appendix 2) but one may check at once that $\hat{\cal{D}}$ and $\hat{\cal{D}}_2$ are completely different operators while the operator $\hat{\cal{D}}_1$ is far from being self-adjoint even though it is described by a $5 \times 5$ operator matrix. Our purpose is to prove that it can be nevertheless transformed in a very tricky way to a self-adjoint operator, exactly like the $3\times 3$ {\it curl} operator in $3$-dimensional classical geometry because $ad(grad)= - div$. It does not seem that these results are known today.  \\
The starting point is the $3\times 5$ first order operator matrix defining the conformal Killing operator $\hat{\cal{D}}$, namely:  \\
\[   \left( \begin{array}{rrr}
\frac{4}{3} d_1 & - \frac{2}{3} d_2 & - \frac{2}{3}d_3  \\
d_2 & d_1 & 0 \\
d_3 & 0 & d_1   \\
- \frac{2}{3}d_1 & \frac{4}{3}d_2 &  - \frac{2}{3}  d_3 \\
0 & d_3 & d_2 
\end{array} \right )  \]
Substracting the fourth row from the first row and mutiplying the fourth row by $\frac{3}{2}$, we obtain the operator matrix:  \\  
\[   \left ( \begin{array}{rrr}
2 d_1 & - 2 d_2 & 0  \\
d_2 & d_1 & 0 \\
d_3 & 0 & d_1   \\
- d_1& 2 d_2 &  -  d_3 \\
0 & d_3 & d_2 
\end{array} \right)  \]
Adding the fourth row to the first, we obtain the operator matrix:  \\
\[   \left ( \begin{array}{rrr}
 d_1 & 0 & - d_3  \\
d_2 & d_1 & 0 \\
d_3 & 0 & d_1   \\
- d_1 & 2 d_2 &  -   d_3 \\
0 & d_3 & d_2 
\end{array} \right)  \]
Adding the first row to the fourth row and dividing by $2$, we obtain the operator matrix:  \\
\[   \left ( \begin{array}{rrr}
 d_1 & 0 & - d_3  \\
d_2 & d_1 & 0 \\
d_3 & 0 & d_1   \\
0 &  d_2 &  -   d_3 \\
0 & d_3 & d_2 
\end{array} \right)  \]
Multiplying the second, fourth and fifth row by $-1$, then multilying the central column of the matrix thus obtained by $-1$, we finally obtain the operator matrix ${\cal{D}}'$:  \\
\[  \left ( \begin{array}{rrr}
 d_1 & 0 & - d_3  \\
- d_2 & d_1 & 0 \\
d_3 & 0 & d_1   \\
0 & d_2 &     d_3 \\
0 & d_3 & - d_2 
\end{array} \right )  \]
We now care about transforming $\hat{\cal{D}}_2$ given in ([16], p 158) by the $5\times 3$ operator matrix:  \\
\[  \left ( \begin{array}{rrrrr}
 - 2 d_3 & 0 &  d_1 & - 2 d_3 & - d_2  \\
2 d_1 & - d_2 & d_3 & 0 & 0  \\
0 & d_1 & 0 & - 2 d_2 & d_3 
\end{array} \right )  \]
Dividing the first column by $2$ and the fourth column by $- 2$, then using the central row as a new top row while using the former top row as new bottom row, we obtain the operator matrix ${\cal{D}}'_2$:  \\
\[   \left ( \begin{array}{rrrrr}
 d_1 & - d_2 & d_3 & 0 & 0  \\
0 & d_1 & 0 &  d_2 & d_3 \\
- d_3 & 0 &  d_1 &  d_3 & - d_2  
\end{array}  \right )   \]
and check that $ad(\hat{\cal{D}}'_2)= - \hat{\cal{D}}'$ like in the Poincar\'{e} sequence for $n=3$ where $ad(div)=-grad$. As the new corresponding operator $\hat{\cal{D}}'_1$ is homogeneous and of order $3$ ({\it care}), we obtain locally $ad(\hat{\cal{D}}'_1)=\hat{\cal{D}}'_1$, a result not evident at first sight (Compare to [16], p 157). \\
The combination of this example with the results announced in ([22]) brings the need to revisit almost entirely the whole conformal geometry in arbitrary dimension and we notice the essential role performed by the Spencer $\delta$-cohomology in this new framework. \\   \\

\noindent
{\bf  5) CONCLUSION}  \\

First of all, we may summarize the results previously obtained by saying that "{\it Janet and Spencer play at see-saw} " because we have the formula $dim(C_r) + dim(F_r) = dim(C_r(E)) $ and the sum thus only depends on $(n,m,q)$ with $n=dim(X), m=dim(E)$ and $q$ is the order of the involutive operator allowing to construct the sequences, but not on the underlying Lie group or Lie pseudogroup group when $E=T$. Hence, the smaller is the background group, the smaller are the dimensions of the Spencer bundles and the higher are the dimensions of the Janet bundles. As a byproduct, we claim that the only solution for escaping is to increase the dimension of the Lie group involved, adding successively $1$ dilatation and $4$ elations in order to deal with the conformal group of space-time while using the Spencer sequence instead of the Janet sequence. In particular, the Ricci tensor only depends on the elations of the conformal group of space-time in the Spencer sequence where the perturbation of the metric tensor does not appear any longer contrary to the Janet sequence. It finally follows that Einstein equations are not mathematically coherent with group theory and  formal integrability. In other papers and books, we have also proved that they were also not coherent with differential homological algebra which is providing intrinsic properties as the extension modules do not depend on the sequence used for their definition, a quite beautiful but difficult theorem indeed. The main problem left is thus to find the best sequence and/or the best group that must be considered. Presently, we hope to have convinced the reader that only the Spencer sequence is clearly related to the group background and must be used, on the condition to change the group. As a byproduct, we may thus finally say that the situation will not evolve in GR as long as people will not acknowledge the existence of these new {\it purely mathematical} tools and their {\it purely mathematical} consequences. Summarizing this paper in a few words, we do believe that " God used group theory rather than computer algebra when He created the World "!. \\
 
\newpage

\noindent
{\bf REFERENCES}  \\
\noindent
[1] Aksteiner, S., Andersson L., Backdahl, T., Khavkine, I., Whiting, B.: Compatibility Complex for Black Hole Spacetimes, https://arxiv.org/abs/1910.08756 .  \\
\noindent
[2] Aksteiner, S., Backdahl, T: New Identities for Linearized Gravity on the Kerr Spacetime,Ó Phys. Rev. D 99, 044043 (2019), https://arxiv.org/abs/1601.06084 . \\
\noindent
[3] Aksteiner, S., Backdahl, T.: All Local Gauge Invariants for Perturbations of the Kerr Spacetime, Physical Review Letters 121, 051104 (2018), 
https://arxiv.org/abs/1803.05341 . \\
\noindent
[4] Andersson L., Backdahl, T., Blue, P., Ma, S.: Stability for Linearized Gravity on the Kerr Spacetime,Ó (2019), https://arxiv.org/abs/1903.03859 .  \\
\noindent
[5] Goldschmidt, H.: Prolongations of Linear Partial Differential Equations: I Inhomogeneous equations, Ann. Scient. Ec. Norm. Sup., 4, 1 (1968) 617-625.  \\
\noindent
[6] Khavkine, I.: The Calabi Complex and Killing Sheaf Cohomology, J. Geom. Phys., 113 (2017) 131-169. \\
\noindent
[7] Pommaret, J.-F.: Systems of Partial Differential Equations and Lie Pseudogroups, Gordon and Breach, New York (1978); Russian translation: MIR, Moscow, 1983.\\
\noindent
[8] Pommaret, J.-F.: Lie Pseudogroups and Mechanics, Gordon and Breach, New York, 1988.\\
\noindent
[9] Pommaret, J.-F.: Partial Differential Equations and Group Theory, Kluwer, 1994.\\
\noindent
https://doi.org/10.1007/978-94-017-2539-2    \\
\noindent
[10] Pommaret, J.-F.: Partial Differential Control Theory, Kluwer, Dordrecht, 2001.\\
\noindent
[11] Pommaret, J.-F.: Algebraic Analysis of Control Systems Defined by Partial Differential Equations, in "Advanced Topics in Control Systems Theory", Springer, Lecture Notes in Control and Information Sciences 311, 2005, Chapter 5, pp. 155-223.\\
\noindent
[12] Pommaret, J.-F.: Parametrization of Cosserat Equations, Acta Mechanica, 215 (2010) 43-55.\\
\noindent
https://doi.org/10.1007/s00707-010-0292-y  \\
\noindent
[13] Pommaret, J.-F.: The Mathematical Foundations of General Relativity Revisited, Journal of Modern Physics, 4 (2013) 223-239. \\
\noindent
 https://doi.org/10.4236/jmp.2013.48A022   \\
 \noindent
[14] Pommaret, J.-F.: Relative Parametrization of Linear Multidimensional Systems, Multidim. Syst. Sign. Process., 26 (2015) 405-437.  \\
\noindent
DOI 10.1007/s11045-013-0265-0   \\
\noindent
[15] Pommaret, J.-F.: Airy, Beltrami, Maxwell, Einstein and Lanczos Potentials revisited, Journal of Modern Physics, 7 (2016) 699-728. \\
\noindent
https://doi.org/10.4236/jmp.2016.77068   \\
\noindent
[16] Pommaret, J.-F.: Deformation Theory of Algebraic and Geometric Structures, Lambert Academic Publisher (LAP), Saarbrucken, Germany, 2016. A short summary can be found in "Topics in Invariant Theory ", S\'{e}minaire P. Dubreil/M.-P. Malliavin, Springer Lecture Notes in Mathematics, 1478 (1990) 244-254.\\
\noindent
https://arxiv.org/abs/1207.1964  \\
\noindent
[17] Pommaret, J.-F.: Why Gravitational Waves Cannot Exist, J. of Modern Physics, 8 (2017) 2122-2158.  \\
\noindent
https://doi.org/104236/jmp.2017.813130    \\
\noindent
[18] Pommaret, J.-F.: New Mathematical Methods for Physics, Mathematical Physics Books, Nova Science Publishers, New York, 2018, 150 pp. \\
\noindent
[19] Pommaret, J.-F.: Minkowski, Schwarzschild and Kerr Metrics Revisited, Journal of Modern Physics, 9 (2018) 1970-2007.  \\
\noindent
https://doi.org/10.4236/jmp.2018.910125  (arXiv:1805.11958v2 ).  \\
\noindent
[20] Pommaret, J.-F.: Generating Compatibility Conditions and General Relativity, J. of Modern Physics, 10, 3 (2019) 371-401.  \\
\noindent
https://doi.org/10.4236/jmp.2019.103025   \\
\noindent
[21] Pommaret, J.-F.: Differential Homological Algebra and General Relativity, J. of Modern Physics, 10 (2019) 1454-1486. \\
\noindent
https://doi.org/10.4236/jmp.2019.1012097   \\
\noindent
[22] Pommaret, J.-F.: The Conformal Group Revisited, https://arxiv.org/abs/2006.03449 .  \\
\noindent
[23] Pommaret, J.-F.: Nonlinear Conformal Electromagnetism and Gravitation, \\
\noindent
https://arxiv.org/abs/2007.01710 .  \\

\end{document}